\documentclass[pre,twocolumn,superscriptaddress,aps,nofootinbib]{revtex4-1}
\usepackage{empheq}
\usepackage{amsmath}
\usepackage{tabularx}
\usepackage{dsfont}
\usepackage{graphicx}
\usepackage[hidelinks]{hyperref}
\usepackage{stackrel}
\usepackage{amssymb,latexsym,mathrsfs}
\usepackage{amsfonts}
\usepackage{srcltx}
\usepackage{color}
\usepackage{cancel}
\usepackage{xr}
\usepackage{placeins}

\def\d{{\rm d}}
\newcommand{\la}{\langle}
\newcommand{\ra}{\rangle}

\allowdisplaybreaks

\newcolumntype{Y}{>{\centering\arraybackslash}X}
\newcolumntype{L}[1]{>{\raggedright\let\newline\\\arraybackslash\hspace{0pt}}m{#1}}
\newcolumntype{C}[1]{>{\centering\let\newline\\\arraybackslash\hspace{0pt}}m{#1}}
\newcolumntype{R}[1]{>{\raggedleft\let\newline\\\arraybackslash\hspace{0pt}}m{#1}}

\newcommand*{\whitedagger}{\color{white}^\dagger}

\def\0{\emptyset}

\begin{document}
	
\title{Two-dimensional Ising model on random lattices with constant coordination number}
\author{Manuel Schrauth}
\author{Julian A. J. Richter}
\affiliation{Institute of Theoretical Physics and Astrophysics,	University of W\"urzburg, 97074 W\"urzburg, Germany}

\author{Jefferson S. E. Portela}
\affiliation{Institute of Theoretical Physics and Astrophysics,	University of W\"urzburg, 97074 W\"urzburg, Germany}
\affiliation{Departamento Acad\^emico de F\'{\i}sica, Universidade Tecnol{\'o}gica Federal do Paran{\'a}, Pato Branco, 85503-390, PR, Brazil}

\begin{abstract}
	We study the two-dimensional Ising model on a network with a novel type of quenched topological (connectivity) disorder. We construct random lattices of constant coordination number and perform large scale Monte Carlo simulations in order to obtain critical exponents using finite-size scaling relations. We find disorder-dependent effective critical exponents, similar to diluted models, showing thus no clear universal behavior.
	Considering the very recent results for the two-dimensional Ising model on proximity graphs and the coordination number correlation analysis suggested by Barghathi and Vojta (2014), our results indicate that the planarity and connectedness of the lattice play an important role on deciding whether the phase transition is stable against quenched topological disorder.
\end{abstract}
\pacs{PACS numbers: 64.60.De, 75.10.Hk, 05.50.+q, }
\maketitle

\def\aj{AJ}%
\def\araa{ARA\&A}%
\def\arfm{ARFM}%
\def\apj{ApJ}%
\def\apjl{ApJ}%
\def\apjs{ApJS}%
\def\ao{Appl.~Opt.}%
\def\apss{Ap\&SS}%
\def\aap{A\&A}%
\def\aapr{A\&A~Rev.}%
\def\aaps{A\&AS}%
\def\azh{AZh}%
\def\baas{BAAS}%
\def\jrasc{JRASC}%
\def\memras{MmRAS}%
\def\mnras{MNRAS}%
\def\pra{Phys.~Rev.~A}%
\def\prb{Phys.~Rev.~B}%
\def\prc{Phys.~Rev.~C}%
\def\prd{Phys.~Rev.~D}%
\def\pre{Phys.~Rev.~E}%
\def\prl{Phys.~Rev.~Lett.}%
\def\pasp{PASP}%
\def\pasj{PASJ}%
\def\qjras{QJRAS}%
\def\skytel{S\&T}%
\def\solphys{Sol.~Phys.}%
\def\sovast{Soviet~Ast.}%
\def\ssr{Space~Sci.~Rev.}%
\def\zap{ZAp}%
\def\nat{Nature}%
\def\iaucirc{IAU~Circ.}%
\def\aplett{Astrophys.~Lett.}%
\def\apspr{Astrophys.~Space~Phys.~Res.}%
\def\bain{Bull.~Astron.~Inst.~Netherlands}%
\def\fcp{Fund.~Cosmic~Phys.}%
\def\gca{Geochim.~Cosmochim.~Acta}%
\def\grl{Geophys.~Res.~Lett.}%
\def\jcp{J.~Chem.~Phys.}%
\def\jgr{J.~Geophys.~Res.}%
\def\jqsrt{J.~Quant.~Spec.~Radiat.~Transf.}%
\def\memsai{Mem.~Soc.~Astron.~Italiana}%
\def\nphysa{Nucl.~Phys.~A}%
\def\physrep{Phys.~Rep.}%
\def\physscr{Phys.~Scr}%
\def\planss{Planet.~Space~Sci.}%
\def\procspie{Proc.~SPIE}%
\def\jpa{J.~Phys.~A: Math.~Theor.}%
\def\prep{Phys.~Rep.}%
\def\rmp{Rev.~Mod.~Phys.}%
\def\ptps{Prog.~Theor.~Phys.~Supp.}%
\def\cpc{Comput.~Phys.~Commun.}%
\def\jstate{J.~Stat.~Mech.-Theory~E.}
\def\epja{Eur.~Phys.~J.~A}%
\def\epjb{Eur.~Phys.~J.~B}%
\def\syszool{Sys.~Zool.}%
\def\jpcm{J.~Phys.-Condens.~Mat.}
\def\jpssp{J.~Phys.~Part~C~Solid}%
\def\jstat{J.~Stat.~Phys.}%
\def\zfpa{Z.~Phys.~A-Hadron.~Nucl.}
\def\plb{Phys.~Lett.~B}
\def\ijmpc{Int.~J.~Mod.~Phys.~C}
\def\fundmath{Fund.~Math.}
\def\cgp{Comput.~Graph.~Forum}
\def\physicaA{Physica A}
\def\jpcs{J.~Phys.~Conf.~Ser.}
\def\npb{Nucl.~Phys.~B}
\def\jcp{J.~Chem.~Phys.}
\def\dcg{Lect.~Notes~Comput.~Sc.}
\def\scirep{Sci.~Rep.}
\def\epjplus{Eur.~Phys.~J.~Plus}
\def\ptp{Prog.~Theor.~Phys.}%
\def\dam{Discrete~Appl.~Math.}
\def\jpcm{J.~Phys.-Condens.~Mat.}

\let\astap=\aap
\let\apjlett=\apjl
\let\apjsupp=\apjs
\let\applopt=\ao

\section{Introduction}
	Magnetic phase transitions have attracted considerable attention over the last decades in both theory \cite{nishimori2010,zinn2007,mccoy2014} and experiment \cite{binder1986,belanger2000,wildes2017}. 
	For real materials, non-magnetic impurities and structural defects can be important and are modeled through lattice irregularities. In discrete settings, the location of the critical point is non-universal and depends on the coordination number $q$.
	For example, $T_c\approx2.269$ for the 2D Ising model~\cite{ising1925} with nearest-neighbor interactions on a square lattice ($q=4$), $T_c\approx3.641$ for a triangular lattice ($q=6$) and $T_c\approx1.519$ for the honeycomb lattice ($q=3$), see, e.g., \cite{fisher1967}. However, if we consider a \emph{random} lattice, the coordination number usually varies from site to site. Prominent examples are site- and bond-diluted regular lattices~\cite{selke1994}, as well as triangulations of Poissonian point clouds (e.g., of the Voronoi-Delaunay kind~\cite{janke1993,janke1994}). In the case of random dilution, disorder is generic in the sense that it is introduced in a completely uncorrelated manner, since sites or bonds are independently removed according to a given probability. Triangulations and other tilings, as well as general proximity graphs~\cite{tamassia2013}, in contrast, are subject to geometrical constraints, and fall under the term \emph{topological disorder}~\cite{janke2004}.
	
	All those types of random structures show fluctuations in their local degree or coordination number\footnote{In this article we occasionally use network terminology. In particular, \emph{coordination number} and \emph{vertex degree} are used synonymously. The same holds for \emph{lattice} and \emph{network/graph}, and for \emph{link/bond/connection}.}. This, in turn, leads to a  different individual transition temperature for each correlation volume $\xi^d$, i.e., a distribution of $T_i$'s centered on the average critical temperature, $T_c$, instead of one sharp transition point. The width $\Delta T_i$ of the resulting distribution is known to be the crucial quantity that determines whether the transition is stable against disorder~\cite{harris1974}. More specifically, $\Delta T_i$, which measures the fluctuation in the local distance from criticality is compared to $|T-T_c|$, where $T$ denotes the simulation (or experimental) temperature. If $\Delta T_i<|T-T_c|$ is fulfilled as $T\to T_c$, the transition is stable. This famous result can be expressed as a simple inequality, $d\nu>2$, known as the Harris criterion~\cite{harris1974, harris2016}. Here, $\nu$ is the critical exponent of the correlation length $\xi$ and $d$ denotes the dimension of the system.
	
	It has been shown in \cite{barghathi2014} that second order phase transitions on random Voronoi-Delaunay lattices are characterized by a modified Harris criterion, which is explained in terms of strong spatial anti-correlations in coordination numbers. Interestingly, we observe that removing some of the bonds of a Voronoi-Delaunay lattice, as prescribed for obtaining the Gabriel graph~\cite{gabriel1969}, eliminates the anti-correlations. This seems rather puzzling, given the fact that Schawe \emph{et~al.}~very recently found strong evidence that the 2D Ising universality is preserved for those lattices~\cite{schawe2017}, providing an indication that universal properties are not solely dictated by anti-correlations in coordination numbers.
	
	We propose a novel random lattice construction with fixed local coordination number in order to suppress fluctuations in the local transition temperature. We call the model \emph{Constant Coordination} (CC). The remaining fluctuations of $T_i$ among the independent disorder realizations, which are revealed in our Monte Carlo simulations, can therefore only originate from the implicit connectivity disorder, as there are no degree fluctuations by construction. In order to clarify the question whether this kind of topological disorder renders the two-dimensional Ising transition unstable, we perform large scale Monte Carlo simulations, calculate the critical exponents of several observables and compare them to their corresponding universal values.
	
	The structure of the paper is the following. A short review of various network topologies and a presentation of the geometric aspects and algorithmic details of the CC random lattice is given in Sec.~\ref{sec:LatticeModels}. In Sec.~\ref{sec:Model_and_Methods} we summarize the Monte Carlo methods employed for the two-dimensional Ising model, including the calculation of observables and details of how the quenched averages are performed. The results of our simulations are presented in Sec.~\ref{sec:Results}, followed by a discussion in the context of other types of quenched disorder, namely Voronoi-Delaunay triangulations and diluted lattices, in Sec.~\ref{sec:Discussion}. Finally, Sec.~\ref{sec:Conclusion} summarizes our findings.
\section{Lattice Models}
\label{sec:LatticeModels}

	The behavior of the Ising model is determined both by its Hamiltonian (see Sec.~\ref{sec:Model_and_Methods}) and by the network topology, or lattice structure, which describes how the sites are linked to each other. The number of nearest neighbors of a given site, which is the number of sites it is directly connected to, is the coordination number or degree $q$ of the site. We denote by $N$ the number of lattice sites and by $L=N^{1/d}$ its linear size, a quantity appropriate to state results for an arbitrary dimension $d$. In this work we consider only lattices on a torus, i.e., in $d=2$ dimensions with periodic boundary conditions.

\subsection{Coordination Number Fluctuations}

	It is shown by Barghathi and Vojta in \cite{barghathi2014} that coordination number fluctuations in random lattices play a crucial role in determining the effect of disorder on phase transitions. In their work, the scaling of disorder fluctuations with increasing length scale is used to determine whether the considered type of disorder is capable of altering the critical exponents at the transition. Specifically, a two-dimensional random lattice of size $N = L^2$ is partitioned into $N_b$ blocks of size $L_b^2$, where the average coordination number within each block $\mu$ is given by
	\begin{align}
	\label{eq:blockingAnalysis1}
	Q_\mu = \frac{1}{N_\mu} \sum \limits_{i=1}^{N_\mu} q_i.
	\end{align}
	Here, $N_\mu$ denotes the number of lattice sites contained in block $\mu$ and $q_i$ is the coordination number of the site~$i$. The standard deviation of $Q_\mu$, which is used to quantify coordination number fluctuations, reads
	\begin{align}
	\sigma_Q(L_b) &= \sqrt{ \frac{1}{(L/L_b)^2 - 1} \sum \limits_{\mu=1}^{(L/L_b)^2} \big( Q_\mu - \bar{q} \big)^2   }, \label{eq:flucScale1}
	\end{align}
	where $\bar{q}$ denotes the \emph{asymptotic} average coordination number of the lattice and we use that $N_b=L^2/L_b^2$. These disorder fluctuations can then be investigated on different length scales by evaluating Eq.~\ref{eq:flucScale1} for different $L_b$. Fig.~\ref{fig:blocking} shows $\sigma_Q(L_b)$ for various lattice models, which are described in the following sections. 
	
	In order to measure anti-correlation effects quantitatively, we calculated the \emph{connected two-point correlation function of the coordination number}~\cite{barghathi2014}
	\begin{align}
	C(\mathbf{r}) = \dfrac{1}{N}\sum_{i,j}(q_i-\bar{q})(q_j-\bar{q})\delta(\mathbf{r}-\mathbf{r}_{ij}).
	\label{eq:corr1}
	\end{align}
	 Here, $\mathbf{r}_{ij}$ denotes the distance vector from site $i$ to $j$. Obviously, $C(\bf{r})$ is identically zero for constant coordination lattices (see Sec.~\ref{sec:CC}). Therefore, we also introduce the \emph{connected two-point correlation function of the second-layer coordination number}, defined by
	\begin{align}
	C^\mathrm{2nd}(\mathbf{r}) = \dfrac{1}{N}\sum_{i,j}(q_i^\mathrm{2nd}-\bar{q}^\mathrm{2nd})(q_j^\mathrm{2nd}-\bar{q}^\mathrm{2nd})\delta(\mathbf{r}-\mathbf{r}_{ij}),
	\label{eq:corr2}
	\end{align}
	 where $q_i^\mathrm{2nd}$ denotes the number of \emph{next-nearest} neighbors, i.e., the number of sites that can be reached from point $i$ by exactly two links and at the same time are not part of the nearest neighbors. This quantity should capture similar geometrical information as its first-layer equivalent, $C(\mathbf{r})$.

\subsection{Voronoi-Delaunay Construction}
\label{sec:VD_lattice}
	
	The Delaunay triangulation for a set of points is a triangulation in which the circumcircle of every triangle is empty, i.e., contains no point of the set. Such triangulations contain as a subgraph the (first) nearest-neighbor graph (see Section~\ref{sec:ProximityGraphs}) and guarantee that the distance along the edges between any two points is not larger than about $2.42$ times their metric distance~\cite{okabe2000}. Regarding edges as neighboring relations, we refer to the Delaunay triangulation spanning a given set of points as the \emph{Voronoi-Delaunay} (VD) lattice of this set.
	An example of such a lattice for a Poissonian sampling is shown in Fig.~\ref{fig:geometry}.
	For computing VD triangulations, we employ the CGAL library~\cite{kruithof2017}.
	
\begin{figure}
		\centering
		\includegraphics[width=\linewidth]{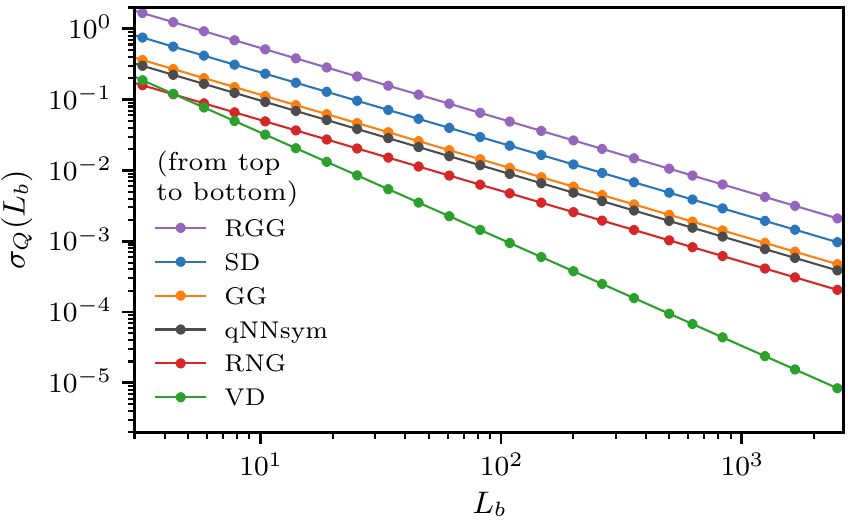}
		\caption{Coordination number fluctuations on different length scales for several lattices. The curves are obtained using Eq.~\ref{eq:flucScale1} and found to decay as $\sigma_Q \sim L_b^{-3/2}$ for the Voronoi-Delaunay triangulation (VD), and as $\sigma_Q \sim L_b^{-1}$ for the others. Measured decay exponent values: Gabriel Graph (GG): $0.999(1)$, Relative Neighborhood Graph (RNG): $1.001(2)$, Site-Diluted regular square lattice (SD): $1.001(3)$, VD: $1.501(2)$, Random Geometric Graph (RGG): $1.004(6)$ and symmetrized $q$-Nearest-Neighbor graph (qNNsym) with $q\geq6$: $1.001(2)$.}
		\label{fig:blocking}
	\end{figure}

	The Ising model has been thoroughly studied on two- and three-dimensional VD lattices (see~\cite{janke1993,janke1994,lima2000} and \cite{janke2002,lima2008}) and found to belong to the same universality class as the pure model, both for constant as well as distance-dependent couplings. Whereas the 2D Ising model represents a marginal case of the Harris criterion $(d\nu=2)$, the unchanged universality in 3D was surprising, since the criterion is violated. This particular result partially motivated the study of coordination number fluctuations in \cite{barghathi2014}. There, using geometric arguments, it was reasoned that the total coordination number in Voronoi-Delaunay lattices with periodic boundary conditions is constant in each instance. This constraint generates anti-correlations in the local coordination number\footnote{For example, a highly connected node will typically be surrounded by less connected nodes.} and it is shown that connectivity disorder in VD lattices decays as fast as $\sigma_Q \sim L_b^{-3/2}$. In contrast to that, in systems with uncorrelated disorder, such as randomly site- or bond-diluted models, $\sigma_Q \sim L_b^{-1}$ holds, as can be seen from Fig.~\ref{fig:blocking}.
	
	\begin{figure}
		\centering
		\includegraphics[width=\linewidth]{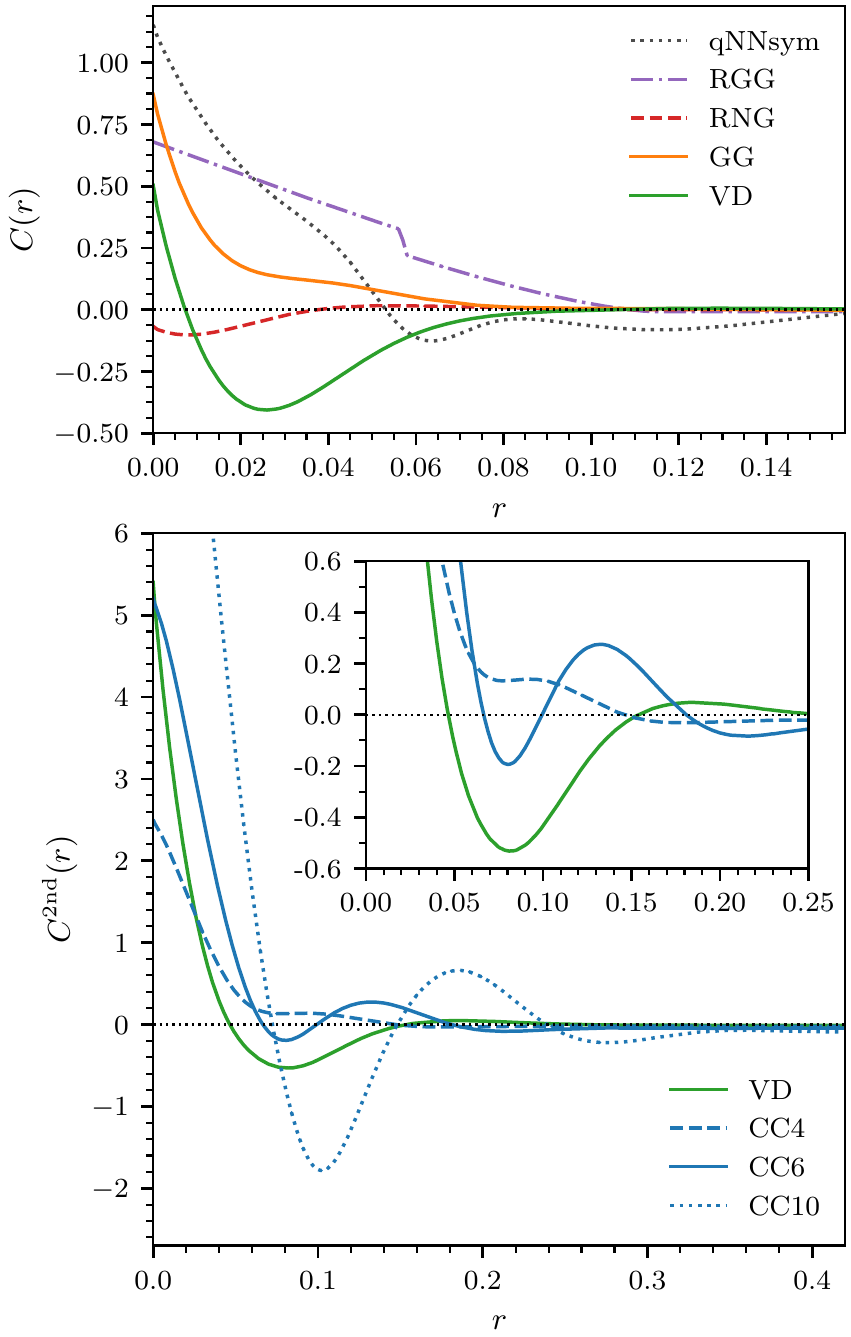}
		\caption{Upper panel: Coordination number correlation function, Eq.~\ref{eq:corr1}, for the Voronoi-Delaunay triangulation, Gabriel Graph, Relative Neighborhood Graph, Random Geometric Graph and symmetrized $q$-Nearest-Neighbor graph with $q\!\geq\!6$. For the RGG the radius was chosen such that the average coordination number is $\bar{q}=6$ and $C(r)$ is rescaled by a factor of $0.1$. Lower panel: Second-layer coordination number correlation function \eqref{eq:corr2} for VD and the constant coordination models (CC4, CC6 and CC10). We show a magnification in the inset.}
		\label{fig:Corr}
	\end{figure}

\subsection{Proximity Graphs}
\label{sec:ProximityGraphs}

	Graphs whose sites lie on a metric space and are connected whenever they are, according to a given criterion, sufficiently close together, are called \emph{proximity graphs}~\cite{tamassia2013}. Different proximity criteria correspond to different graph constructions. One such construction is the VD graph, presented in Sec.~\ref{sec:VD_lattice}. Other proximity graphs we consider are the \emph{Random Geometric Graph} (RGG)~\cite{penrose2003}, the \emph{Gabriel Graph} (GG)~\cite{gabriel1969} and the \emph{Relative Neighborhood Graph} (RNG)~\cite{toussaint1980}. These lattices are described below and can be efficiently calculated for a Euclidean metric~\cite{jaromczyk1992}.
	
\begin{figure*}
		\centering
		\includegraphics[width=0.31\linewidth]{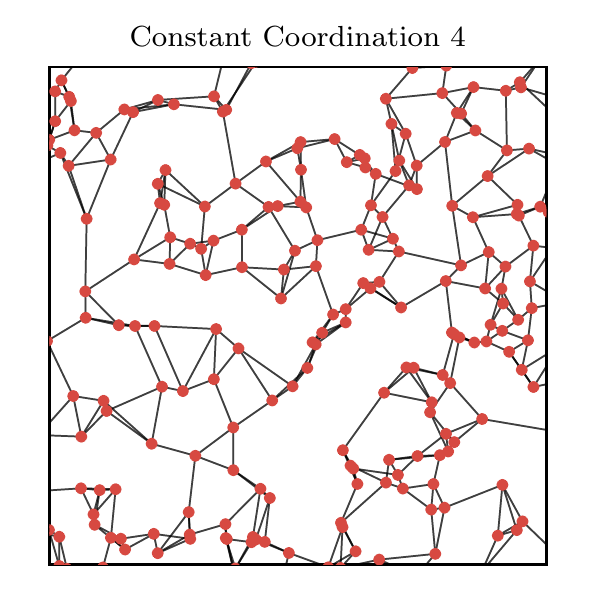}
		\includegraphics[width=0.31\linewidth]{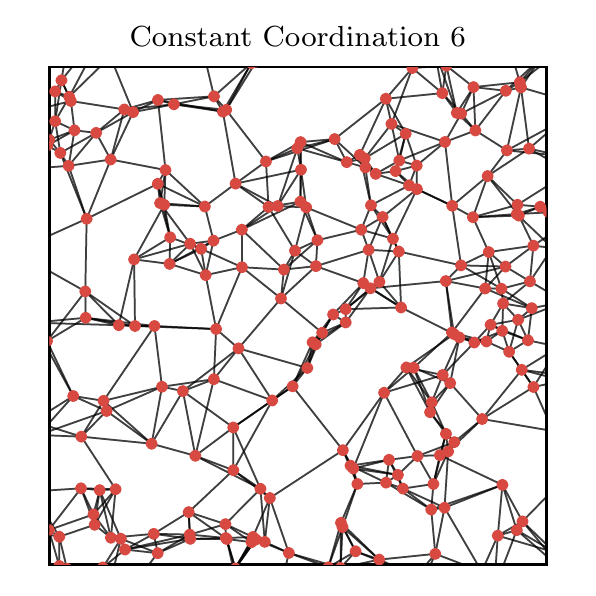}
		\includegraphics[width=0.31\linewidth]{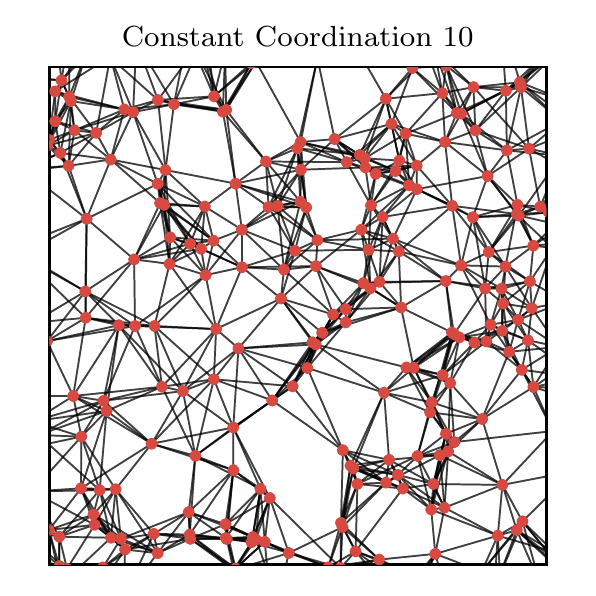}\\
		\includegraphics[width=0.31\linewidth]{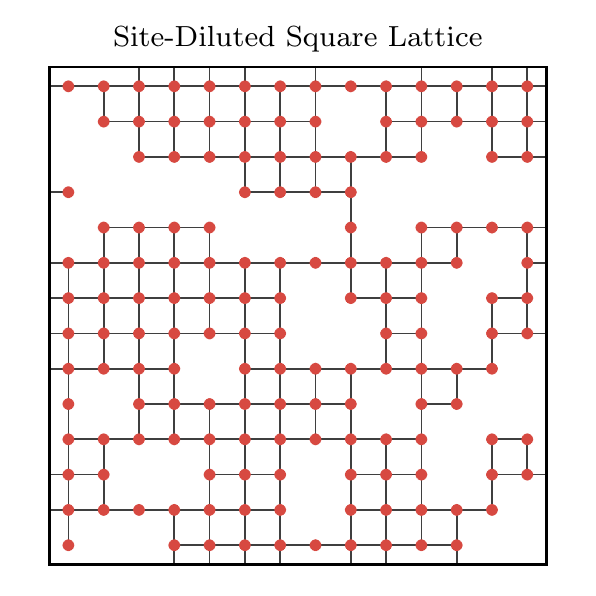}
		\includegraphics[width=0.31\linewidth]{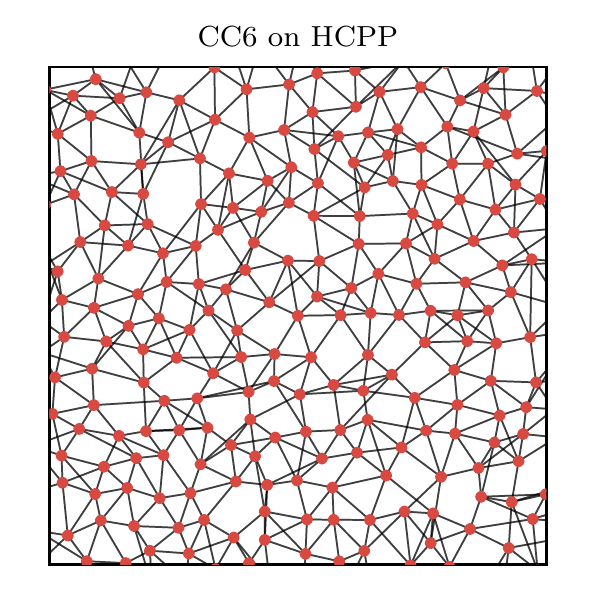}\\
		\includegraphics[width=0.31\linewidth]{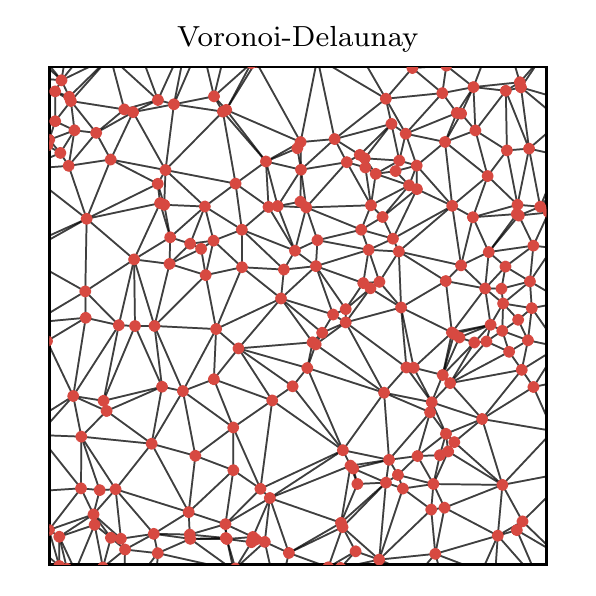}
		\includegraphics[width=0.31\linewidth]{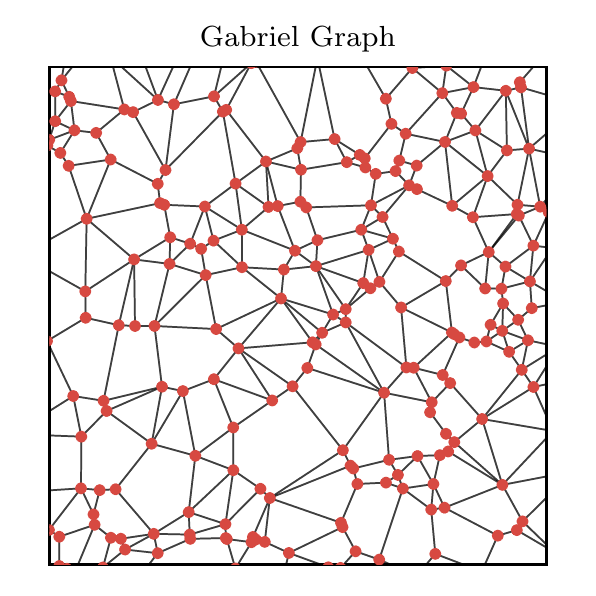}
		\includegraphics[width=0.31\linewidth]{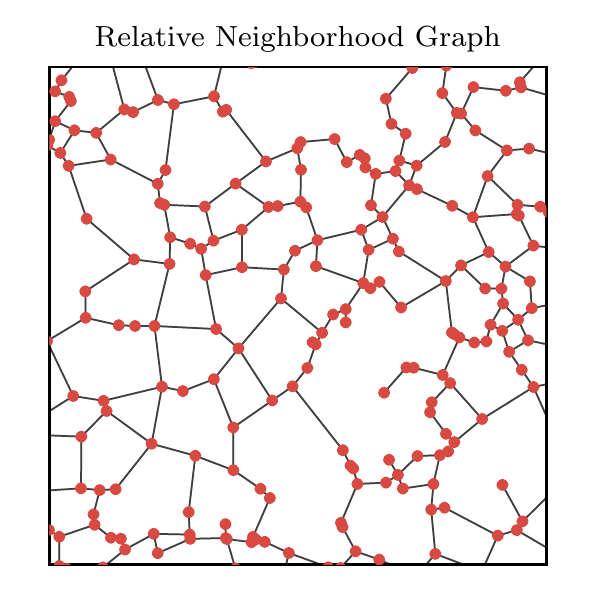}\\
		\includegraphics[width=0.31\linewidth]{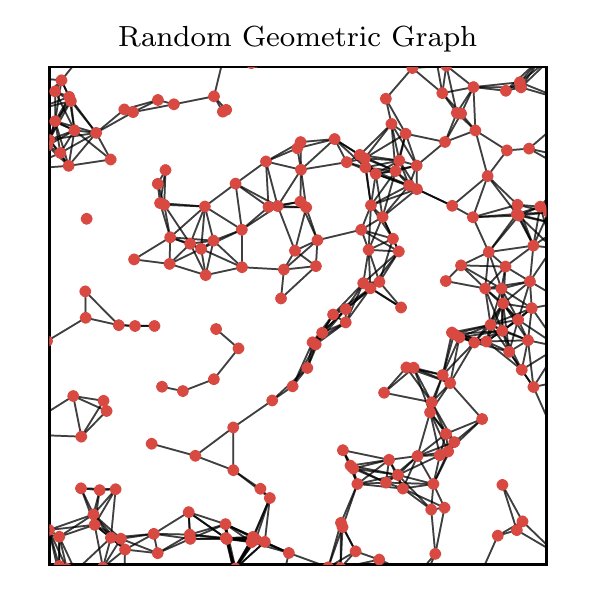}
		\includegraphics[width=0.31\linewidth]{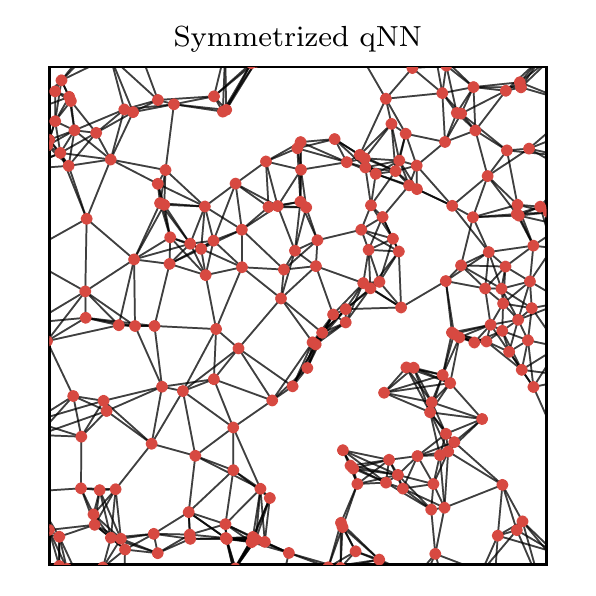}
		\caption{Samples of the lattice constructions considered in this article. Note that all lattices with exception of those in the second line are constructed from the same set of points. For the RGG, the radius is set according to Eq.~\ref{eq:CaptureRadius} such that $\bar{q}=6$. The dilution probability for the diluted lattice is 20\%. For the symmetrized $q$-Nearest-Neighbor lattice, $q\geq6$ holds, as all links have been made bidirectional. } 
		\label{fig:geometry}		
	\end{figure*}

	In an RGG, any two points whose distance falls below a certain threshold are linked. In two dimensions, these graphs can be defined using the auxiliary variable
	\begin{align}
		R_{\bar{q}} = \sqrt{\dfrac{\bar{q}}{\pi L^2}},
		\label{eq:CaptureRadius}
	\end{align}
	which denotes the interaction radius of a random geometric graph with $\bar{q}$ neighbors on average.
	For a comprehensive review see~\cite{barthelemy2011}. In these lattices, correlations arise from the fact that a high degree node must be surrounded by many points close to each other, which typically implies rather high coordination numbers in its immediate surrounding as well. In other words, dense clusters are more likely than in generic random networks. This property can be observed very clearly in the example of an RGG lattice shown in Fig.~\ref{fig:geometry}.
	
	In a GG, also displayed in Fig.~\ref{fig:geometry}, two points $i$ and $j$ are connected whenever $d(i,j)^2 \le d(i,k)^2 + d(k,j)^2$ for any other point $k$ of the graph, where $d(i,j)$ is the distance between $i$ and $j$. This condition translates into requiring that the smallest circle defined by $i$ and $j$ contains no other points.	
	The RNG is similarly defined by the more restrictive condition  $d(i,j) \le \max{[d(i,k), d(k,j)]}$ and also shown in Fig.~\ref{fig:geometry}.

	For these three proximity graphs, we repeat the blocking analysis from \cite{barghathi2014}, using Eqs.~\ref{eq:blockingAnalysis1} and \ref{eq:flucScale1}. Fitting the fluctuations to $\sigma_Q\sim L_b^{-a}$ in Fig.~\ref{fig:blocking}, we find decay exponents consistent with $a=1$, which correspond to that of conventional, uncorrelated disorder. This is somewhat unexpected, especially in light of the very recent results from \cite{schawe2017}, which provided unambiguous evidence that the 2D Ising model on the RNG and GG falls into the universality class of the regular model. For that reason, we also repeat the calculation of the coordination number correlation function from \cite{barghathi2014}, in order to shed light on the role of anti-correlations in the coordination number. The results are shown in Fig.~\ref{fig:Corr}, compared to VD and random geometric graphs. Interestingly, the curve for the GG remains positive, i.e., it displays no anti-correlation at all and is thus consistent with the slow disorder decay observed above. It is remarkable that the pruning of bonds of a VD lattice in order to obtain the GG causes such a significant change with respect to the coordination number correlations. Equally surprising is the circumstance that the removal of further bonds from the GG, leading to the RNG, results in negative correlations for short ranges. That means that highly connected sites tend to be linked to less connected sites, and vice versa. The RGG curve reflects the high clustering mentioned above, falling linearly up to the interaction radius, $R_{\bar{q}}$, where it displays a pronounced drop before approaching zero for distances around $r=2 R_{\bar{q}}$. This is consistent with the fact that, for two sites with non-overlapping interaction regions, the coordination numbers are effectively uncorrelated.
	
\subsection{Constant Coordination Lattice}
\label{sec:CC}	
	
	In this work, our aim is to obtain a random lattice for which the local coordination number is constant for all points by construction. Since the local coordination number does not fluctuate, $\sigma_Q$ -- of course -- vanishes on any length scale. Furthermore, the constraint of a constant \emph{total} coordination number is also trivially fulfilled.
	
	When imposing the constraint mentioned above, the perhaps most obvious lattice construction one can think of would be a $q$-Nearest-Neighbor lattice, where every site is linked to the $q$ spatially closest sites. This construction is very simple (e.g., compared to VD), since no geometrical information other than the point distances is required, and is straightforward in any dimension. However, this lattice is in general a \emph{directed graph}, since neighborhood is not necessarily reciprocal. Therefore, in the resulting lattice, only $q_\mathrm{out}$, the out-degree of every site, is constant, i.e., exactly $q$ bonds emanate from each site. Since not all links are bi-directional, though, $q_\mathrm{in}\neq \mathrm{const.} $
	
	In the past, it has been pointed out by several authors that equilibrium systems on directed graphs can be regarded as pathological in the sense that the detailed balance condition is violated \cite{godreche2009}. This leads to the fact that, e.g., on a directed, scale-free Barab\'asi-Albert graph, no spontaneous magnetization can be found and different update algorithms give different results~\cite{lima2006}. On directed small-world networks, the $S=1$, $3/2$ and 2 Ising model, as well as the Blume-Capel model, show a phase transition which changes from second to first order if a specific critical rewiring probability is exceeded \cite{lima2009,fernandes2010,lima2013}. In the second-order regime, the aforementioned results indicate a different universality class compared to the corresponding models on a regular lattice. For a recent review article, see also \cite{lima2014}. Although those results have been calculated using traditional equilibrium Monte Carlo simulations, it was first pointed out in \cite{sanchez2002} that those directed systems can be seen as being in a non-equilibrium stationary state rather than in conventional equilibrium. Therefore, even a proper definition of the energy of the system becomes problematic~\cite{godreche2009}.
	
	In order to avoid the massive complications accompanied with directedness, there are two common ways to symmetrize $q$-Nearest-Neighbor constructions. One can either delete any directed links, such that only the bi-directed ones are left, or also add the reverse links to the nodes connected by directed ones. Obviously, lattice sites can be left with more than $q$ neighbors after the latter symmetrization procedure, and can have fewer than $q$ neighbors after the former procedure. Additionally, it can easily be checked that either symmetrization does not lead to a  constant global coordination number $ Q_\text{tot} $, which means that $ Q_\text{tot} $ is (slightly) different for each lattice realization. Furthermore, the blocking analysis of for those two possibilities clearly shows a decay as consistent with conventional, uncorrelated disorder, i.e., $\sigma_Q \sim L_b^{-1}$, as expected.
	We also display the correlation function for the symmetrized $q$-Nearest-Neighbors lattice in Fig.~\ref{fig:Corr}.

	We want to construct an undirected lattice model where every site has \emph{exactly} $q$ neighbors. Naively linking every point to some other randomly chosen points that still have fewer than $q$ neighbors would lead to mean-field behavior, similar to small-world networks~\cite{herrero2002} and Erd\"{o}s-R\'{e}nyi graphs \cite{lima2012}, since the mean path length is then of the same order as the system size and therefore information propagates effectively instantaneously through the lattice. We therefore place as a particular demand on our model that the interactions are short-ranged in the sense that the bond lengths $ \ell\ll L $. The resulting construction, which we refer to as \textit{Constant Coordination} (CC) lattice, works as follows.
	
	\emph{Procedure}: We start with the fully random graph mentioned above, where one point at a time is linked to $q$ other points, randomly chosen from those with fewer than $q$ neighbors. Afterwards, the sites are dynamically rewired by a simulated annealing algorithm \cite{kirkpatrick1983}, respecting the constraint of fixed $ q $. More specifically, the algorithm chooses two links, $ \overline{i j} $ and  $ \overline{k l} $, at random and checks whether a rewiring of the connections to $ \overline{i l} $ and $ \overline{j k} $ would lead to a decrease of the sum of the bond lengths, i.e.,
	\begin{align}
		d(i,l)+d(j,k) < d(i,j) +d(k,l).\label{eq:constraint}
	\end{align}
	If this inequality is obeyed, the change is accepted and the algorithm moves on by considering the next pair of links. If instead, the new configuration would lead to an increase of the combined link lengths, the rewiring is accepted only with probability $ \exp(-\Delta H/T) $, where
	\begin{align}
		\Delta H \equiv d(i,j)+d(k,l) - (d(i,l)+d(j,k))
	\end{align}
	defines the cost function. The  non-zero simulated annealing temperature $T$ has the effect of noise on the convergence to a state of low cost function. The value of $T$ is logarithmically decreased during the simulation, such that in the beginning, ``bad" rewiring updates are accepted with moderate probability, whereas in the final stages, this probability almost vanishes. More details of our algorithm can be found in Appendix \ref{sect:sa}. Samples of the lattice is shown in Fig.~\ref{fig:geometry}.

	As the degree fluctuations are trivially equal to zero, Fig.~\ref{fig:Corr} shows also the second-layer degree fluctuations according to Eq.~\ref{eq:corr2} for the particular CC models we consider and compares them to those of the VD triangulation. As can be seen, VD exhibits pronounced anti-correlations in the second-layer coordination number as well. The curve for CC10 is qualitatively similar, but shows significantly stronger oscillations. Comparing CC10 with CC6 and CC4, it can be noticed that the relative strength of anti-correlations decreases as $q$ is decreased. For $q=4$ the first minimum is hardly visible and positive values dominate (see inset of the figure).

	It is also worth considering samplings other than the simple Poissonian, such as the Hard Core Point Process (HCPP), where the random points are placed respecting a minimum distance $R_r$ from each other \cite{baddeley2015}. We briefly address this model in Appendix~\ref{sect:hcpp}. A sample of the CC neighbor construction on this hard core point process can be seen in Fig.~\ref{fig:geometry}.
	
\subsection{Link Lengths}

	One of the key ingredients to establish a well-defined magnetic phase transition that does not behave in a mean-field fashion is the locality of interactions, usually realized by establishing nearest- or next-nearest-neighbor couplings on the lattice. In other words, the characteristic interaction range $\ell$ should be small compared to the system dimensions, $\ell\ll L$. As soon as one allows for sufficiently many long-range ``shortcuts", as those found, for instance, on small-world lattices \cite{herrero2002}, the behavior of the system is governed by its mean-field fixed point.
	
\begin{figure}
		\centering
		\includegraphics{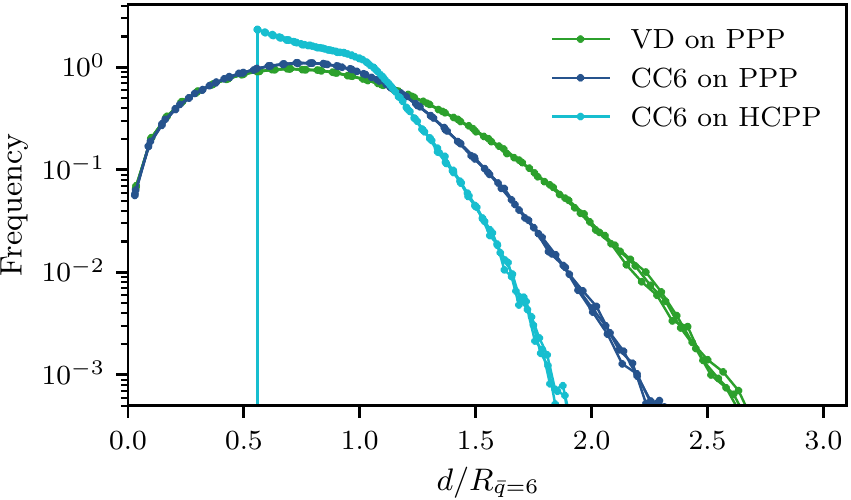}
		\caption{Normalized bond lengths histogram, rescaled in units of $R_{\bar{q}=6}$ for the VD triangulation and the CC6 lattices on Poissonian and Poissonian disc-sampled (see Appendix~\ref{sect:hcpp}) point distributions. Each color shows collapsing curves for $L=32,64$ and $128$. The mean values of the histograms are 0.819 (VD), 0.767 (CC6) and 0.834 (CC6-HCPP). } 
		\label{fig:histogram}		
	\end{figure}

	For this reason, as detailed in Section~\ref{sec:CC}, the CC lattice is specifically designed to be sufficiently local. This property can be quantitatively characterized by means of the link lengths statistics. Fig.~\ref{fig:histogram} shows the normalized link length histogram for the CC lattice with $q=6$ on a Poisson point process, as well as on the hard core point process, compared to the distribution for a Voronoi-Delaunay triangulation ($\bar{q}=6$). For every model, Fig.~\ref{fig:histogram} contains three separate curves (of the same color) which correspond to lattices with $L=32$, $64$ and $128$. The respective curves collapse when rescaled by $R_{\bar{q}=6} $, defined in \eqref{eq:CaptureRadius}, showing that our algorithm provides the correct scaling for finite systems of different linear dimensions $L$. Moreover, Fig.~\ref{fig:histogram} indicates that the CC lattices are even more local than the Voronoi-Delaunay triangulation.
	
	We want to emphasize that the link distance histogram is a sufficient condition to prove locality for our lattice construction, even though it is not a necessary one. If we, for instance, move lattice points to new randomly chosen locations while keeping all bond connections unchanged, we end up with a completely different link length profile with distances of all length scales. However, the topology of the lattice would not be different than before, as it is solely encoded in the neighbor relations. In this context, the typical shortest path length $\bar{L}$ on the graph can be used as a proper quantity to check locality. For regular lattices as well as triangulations on random point clouds, this distance scales as $\bar{L}\sim N^{1/d}$ where $N$ is the number of nodes and $d$ denotes the dimension of the system. Since for our lattices the geometric bond distances are explicitly minimized during the dynamical algorithm, they display the same scaling. Small-world networks, in contrast, show a mean-field type transition \cite{herrero2002} and are known to scale as $\bar{L}\sim \log N$ \cite{watts1998}. Some scale-free networks, on which the temperature of the ferromagnetic to paramagnetic crossover was found to shift with system size and to ultimately diverge for $N\to\infty$ \cite{herrero2004}, even scale as $\bar{L}\sim\log\log N$ \cite{cohen2003}.  
	
\section{Physical Model and Methods}
\label{sec:Model_and_Methods}

	We focus on the two-dimensional, ferromagnetic Ising model, described by the Hamiltonian
	\begin{align}
		\mathcal{H} = -J \sum \limits_{ \langle i,j \rangle} s_i s_j,
		\label{eq:Hamiltonian}
	\end{align}
	where $J>0$ quantifies the coupling between nearest neighbors $\langle i,j \rangle$ and the spin variables assume values $s_i = \pm 1$. On a spatially disordered lattice, a natural option is to consider a distance-dependent coupling constant $J(r)$. However, we want to focus exclusively on connectivity disorder and therefore set $J=1$ throughout this work, in order to avoid any possible effects of random couplings. Besides, in VD lattices, it is known that distance-dependent coupling constants do not affect the universal properties of the phase transition \cite{lima2000}.
		
	In order to study the Ising model in the vicinity of the critical point, we employ importance-sampling Monte Carlo methods, using single-cluster, as well as local-update algorithms. In particular, we use the algorithm proposed by Wolff \cite{wolff1989}, which significantly reduces the critical slowing down near the critical point and is straightforwardly applicable to disordered lattices.
	
	Keeping track of the full time series of measurements of magnetization and energy during the simulations enables us to calculate all observables of interest by means of single-histogram reweighting techniques \cite{ferrenberg1989,ferrenberg1988}. This way, the observables can be obtained as continuous functions of temperature $\beta$, allowing the extremal points used in the finite- size scaling analysis to be determined with high precision. By estimating the valid reweighting range, as proposed in \cite{janke2008}, we make sure that no systematic errors are introduced in our analysis.
	
	In the investigation of disordered systems, it is necessary to average physical observables over many different, independent disorder realizations, also called \emph{replicas}, of the system.	The so-called quenched averages over $N_r$ replicas are performed at the level of (extensive) observables, rather than at the level of the partition function~\cite{binder1986}. Denoting quenched averages as
	\begin{align}
		\label{eq:arithMean}
		[\mathcal{O}]_\text{avg} \equiv \frac{1}{N_r} \sum \limits_{i=1}^{N_r} \mathcal{O}_i
	\end{align}
	and thermal averages as $\langle ... \rangle $,
	we use the following definitions of magnetization, energy, susceptibility and specific heat:
	\begin{subequations}
	\begin{align}
		m &= [ \la |m| \ra ]_\text{avg}, \\
		e &= [ \la e \ra ]_\text{avg}, \\
		\label{eq:obs_chi}
		\chi &= N \beta [ \la m^2 \ra - \la |m| \ra^2  ]_\text{avg},\\
		\label{eq:bs_cv}
		C &= N \beta^2 [ \la e^2 \ra - \la e \ra^2  ]_\text{avg},
	\end{align}
	\end{subequations}
	as well as the following derivatives
	\begin{subequations}
	\begin{align}
		\frac{\d [ \la m \ra ]_\text{avg} }{\d\beta} &= [ \la |m|e \ra - \la |m|  \ra \la e \ra  ]_\text{avg}, \\
		\frac{ \d \ln[ \la |m| \ra ]_\text{avg} }{\d\beta} &= \Bigg[ \frac{\la |m| e \ra }{\la |m| \ra} - \la e \ra  \Bigg]_\text{avg}, \\
		\frac{ \d \ln[ \la m^2 \ra ]_\text{avg} }{\d\beta} &= \Bigg[ \frac{\la m^2 e \ra }{\la m^2 \ra} - \la e \ra  \Bigg]_\text{avg},\\
		\frac{1}{N}\frac{ \d [U_2]_\text{avg} }{\d\beta} &= \Bigg[  \big(  1 - U_2 \big) \Bigg(\la e \ra - 2  \frac{\la |m|e \ra}{\la |m| \ra} + \frac{\la m^2e\ra }{\la m^2\ra } \Bigg) \Bigg]_\text{avg}, \label{eq:dU2}\\
		\label{eq:obs_dU4db}
		\frac{1}{N}\frac{ \d [U_4]_\text{avg} }{\d\beta} &= \Bigg[  \big(  1 - U_4 \big) \Bigg(\la e \ra - 2  \frac{\la m^2e \ra}{\la m^2 \ra} + \frac{\la m^4e\ra }{\la m^4\ra } \Bigg) \Bigg]_\text{avg},
	\end{align}
	\end{subequations}
	which all exhibit singularities close to the phase transition in the thermodynamic limit.
	In Eqs.~\ref{eq:dU2} and \ref{eq:obs_dU4db}, $U_2$ and $U_4$ denote the second- and fourth-order magnetic cumulants, given by
	\begin{align}
		U_2(\beta) &= \Bigg[ 1 - \frac{ \la m^2 \ra }{ 3 \la |m| \ra^2}  \Bigg]_\text{avg}, \\
		U_4(\beta) &= \Bigg[ 1 - \frac{ \la m^4 \ra }{ 3 \la m^2 \ra^2}  \Bigg]_\text{avg}.
	\end{align}
	Note that the intersection point of the fourth-order magnetic cumulant $U_4(\beta)$ for two different lattice sizes yields an estimate for the critical temperature $\beta_c$.
	
	In a finite system of linear size $L$, it is well known that, near the critical point, the above quantities scale as 
	\begin{subequations}
		\begin{align}
			\label{eq:scaling_m}
			[ \la m \ra ]_\text{avg} &= L^{-\beta/\nu} f_m(x)(1+\ldots), \\
			\label{eq:scaling_chi}
			\chi &= L^{\gamma/\nu} f_\chi(x)( 1 + \ldots ), \\
			C &= C_0 + L^{\alpha/\nu} f_C(x)( 1 + \ldots ), \\
			\frac{\d [ \la m \ra ]_\text{avg} }{\d\beta} &= L^{(1-\beta)/\nu} f_{m'}(x)( 1 + \ldots ), \label{eq:scaling_dmdb}\\
			\frac{ \d \ln[ \la |m| \ra ]_\text{avg} }{\d\beta} &= L^{1/\nu} f_{m,1}(x)( 1 + \ldots ), \\
			\frac{ \d \ln[ \la m^2 \ra ]_\text{avg} }{\d\beta} &= L^{1/\nu} f_{m,2}(x)( 1 + \ldots ), \\
			\frac{ \d [U_2]_\text{avg} }{\d\beta} &= L^{1/\nu} f_{U_2}(x)(1+\ldots), \\
			\frac{ d [U_4]_\text{avg} }{\d\beta} &= L^{1/\nu} f_{U_4}(x)(1+\ldots),\label{eq:scaling_dU4db}
		\end{align}
		\label{eq:scaling_relations}
	\end{subequations}
	where $\alpha$, $\beta$, $\gamma$ and $\nu$ are critical exponents, $C_0$ is the regular part of the specific heat that does not diverge at the critical point and the functions $f$ are universal scaling functions with the argument $x$ given by
	\begin{align}
		x = ( \beta - \beta_c)L^{1/\nu}.
	\end{align}
	These equations describe the finite-size scaling (FSS) behavior of the considered observables to first order. Corrections of higher order to the scaling equations are expected to become irrelevant for large system sizes $L$. 
	
	The time series of measurements is resampled into blocks according to the jackknife method~\cite{efron1994}. This procedure is known to decrease the bias of the estimator of the average, $\overline{ \mathcal{O}^{(B)} }$. Furthermore, for regular lattices, where no replica average is necessary, the error is estimated via	
	\begin{align}
		\sigma_\mathcal{O} = \frac{N_B -1}{N_B} \sum \limits_{i=1}^{N_B} ( \mathcal{O}^{(B)}_i - \overline{ \mathcal{O}^{(B)} } ),
	\end{align}
	where $N_B$ denotes the number of blocks, $\mathcal{O}^{(B)}_i$ indicates the average of an observable $\mathcal{O}$ in block $i$ and $\overline{ \mathcal{O}^{(B)} }$ denotes the average of the $N_B$ individual block-averages. Depending on the number of measurements performed in a simulation, the number of bins should be chosen such that the bin size is large compared to the integrated autocorrelation time, and small compared to the length of the entire sample. In our simulations we use between 100 and 1000 bins.
	
	For the disordered models, however, another average (over replicas) is necessary, as pointed out above. We therefore do not use the jackknife errors of the single curves, but instead calculate the uncertainty of the replica-average via a standard error
	\begin{align}
		\label{eq:err1}
		\sigma_\text{replica}^2 = \frac{1}{N_r(N_r-1)} \sum_{i=1}^{N_r} \big( \langle \mathcal{O} \rangle_i - [ \langle \mathcal{O} \rangle ]_\text{avg}  \big)^2,
	\end{align}
	where $N_r$ denotes the number of replicas. This ensures that both the thermal fluctuations, as well as those among different disorder realizations are properly taken into account~\cite{janke2002}. Note that if we do not discard the individual errors but instead combine them to form a weighted average with associated uncertainty, the fluctuations arising from the different disorder realizations are not correctly accounted for.
	The individual curves are not estimators of the replica-averaged observables, but instead only of their replica-\emph{specific} observables. This means that even if we would perform $n \rightarrow \infty$ measurements for one specific disorder realization, the resulting estimates would not converge to the actual values of the replica-averaged curves.
\section{Results}
\label{sec:Results}

In the following, we present the results of our numerical simulations, which were performed in the department's cluster, taking about $30.000$ CPU-days in total.
\subsection{Regular Square Lattice}
\label{sect:singSquare}
	\begin{figure}[t]
		\centering
		\includegraphics{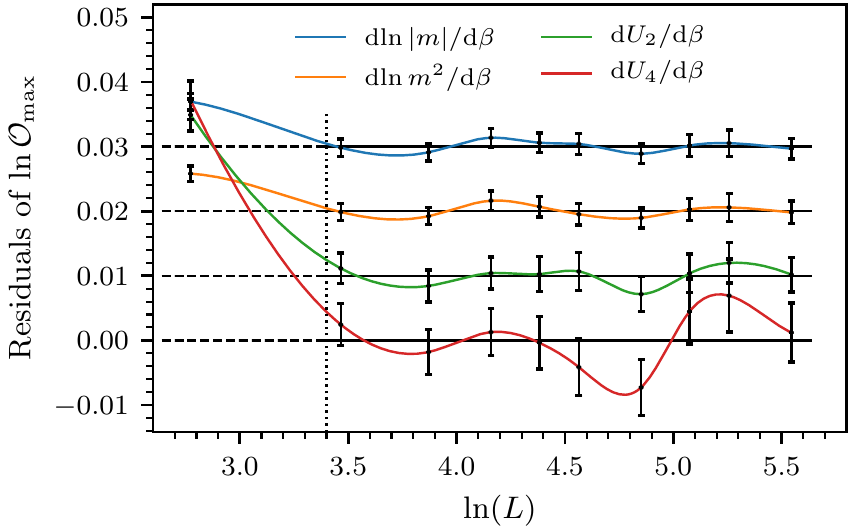}
		\caption{Residuals of 4 out of 28 fits for the exponent $\nu$ for the Ising model on a regular square lattice, shifted vertically for convenience. The vertical black line separates the region that is excluded in the fits. The curves are guides to the eye.} 
		\label{fig:residuals-regular-nu}		
	\end{figure}
	\begin{figure*}[t]
		\centering
		\includegraphics[width=0.495\linewidth]{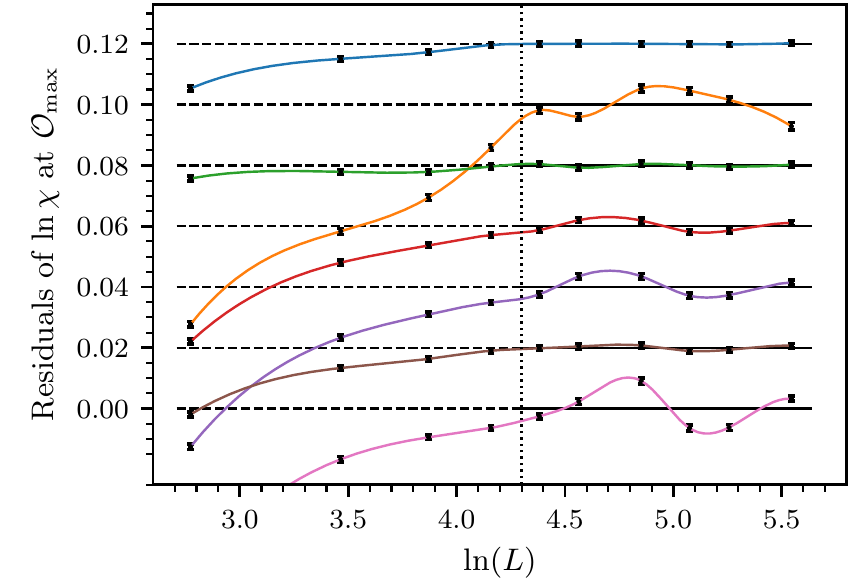}
		\includegraphics[width=0.495\linewidth]{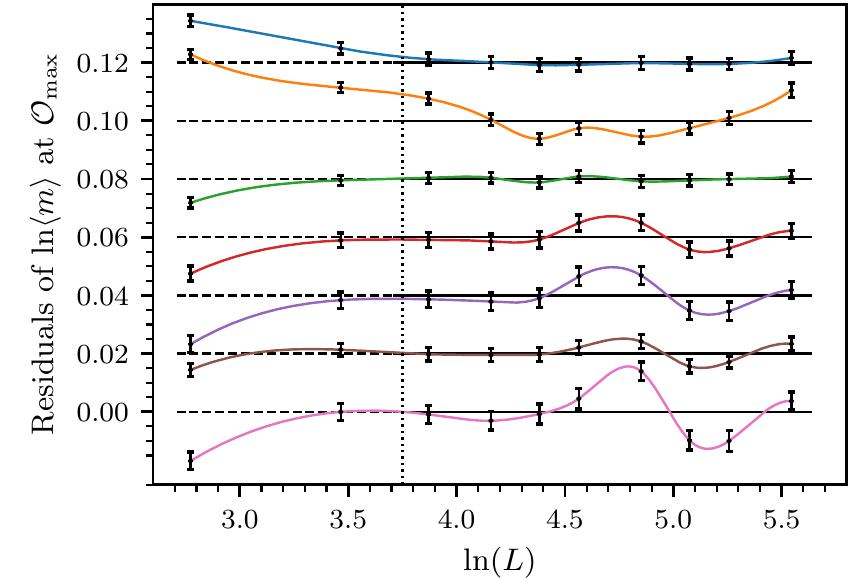}
		\caption{Residuals for the fits of the exponents $\gamma/\nu$ (left panel) and $\beta/\nu$ (right panel) for the regular lattice. The single lines in each panel correspond to observables \eqref{eq:obs_chi} to \eqref{eq:obs_dU4db} from top to bottom and are shifted vertically for convenience. The vertical dotted line separates the region which is excluded from the fitting procedure and the curves are guides to the eye.}
		\label{fig:residuals-regular-gamma-and-beta}		
	\end{figure*}
	As a walk-through of our analysis and validation of our code, we simulate the two-dimensional Ising model, Eq.~\ref{eq:Hamiltonian}, on a regular square lattice with
	$L=16$ to $256$. In total, we perform $7.5\cdot 10^7$ single cluster updates for each system size: the first $2.5\cdot 10^7$ cluster updates are reserved to ensure proper equilibration of the system; after that, magnetization and energy are evaluated every 25th cluster update, yielding $ 2\cdot 10^6 $ almost uncorrelated measurements. Between each measurement, we also perform a full Metropolis sweep \cite{Metropolis1953} in order to make sure that the short-wavelength modes are properly thermalized. In principle, these sweeps are not necessary when using a cluster algorithm on a regular lattice, but for some random lattices, the intermediate Metropolis updates significantly decrease correlations between consecutive configurations. Strongly diluted systems, for example, necessarily require Metropolis sweeps, since cluster updates rarely visit small, isolated components of the lattice.
	As simulation temperatures, we choose the approximate maxima of the susceptibility curves. Reweighting the data returns seven curves \eqref{eq:obs_chi}-\eqref{eq:obs_dU4db} for each system size $ L $; we then determine their maxima, thereby obtaining seven sequences of pseudo-critical temperatures and corresponding function values. Since the system is regular, replica averaging is not necessary here.
	
	The scaling relations \eqref{eq:scaling_relations} generically include multiplicative correction factors of the form $(1+b\cdot L^{-\omega}+\dots)$, with a correction-to-scaling exponent $ \omega $, some non-universal constant $ b $, and possibly further terms of higher order (see, e.g., \cite{kaupuzs2016} for a detailed discussion). 
	Taking into account these corrections would, however, require non-linear fitting methods with at least four parameters, which tend to be numerically unstable.
	In order to avoid non-linear fits while still keeping track of possible systematic corrections, we adopt the following procedure:
    \begin{enumerate}
		\item Determine a suitable minimum lattice size, $L_\mathrm{min}$, by discarding an increasing number of the smallest lattices and refitting, up to the point where the values of the exponents and also the goodness-of-fit parameter $ Q $ \cite{numericalRecipes} cease to show a systematic trend.
		\item Check the corresponding residual plot and, if necessary, increase $L_\mathrm{min}$ in order to eliminate any systematic trend still present in the remaining data points.
    \end{enumerate}
	In order to determine the correlation length exponent $ \nu $, we use the last four scaling relations of \eqref{eq:scaling_relations}, each of which is fitted to the seven pseudo-critical sequences \eqref{eq:obs_chi}-\eqref{eq:obs_dU4db}, yielding a total of 28 fits. The relations could be fitted only at their own pseudo-critical temperatures with good results -- however, performing the full number of fits allows for the determination of $\nu$ to the greatest possible precision. This is advantageous since this exponent is required for the determination of the infinite-volume critical temperature, as well as for the other exponents obtained from $ \gamma/\nu $, $ (1-\beta)/\nu $ and $\beta/\nu$. Nevertheless, we emphasize that taking the full 28 fits into account brings about only a modest increase in precision, given that all fits are trivially correlated, since they stem from the same set of simulations. For the regular lattice, discarding the smallest lattice size simulated, $L=16$, we find 23 acceptable fits with goodness-of-fit values $ Q\geq0.2 $. The residual plots for four of the curves are shown in Fig.~\ref{fig:residuals-regular-nu}. In order to obtain a final value for $\nu$, we calculate the error-weighted average over all acceptable fits. Concerning the uncertainty, we quote the smallest error of the single fits included in the average, thus being quite conservative, as suggested in~\cite{janke2002}. The result is
	\begin{align}
		\nu = 1.0000 \pm 0.0006 \quad (L_\mathrm{min}=32),
	\end{align}
	which perfectly coincides with the analytically known value of $ \nu\!=\!1 $. 
	
	Making use of the relation
	\begin{align}
		\beta_{\mathrm{max}_i} = \beta_c+a_i L^{-1/\nu},
		\label{eq:TcScaling}
	\end{align}
	in combination with the pseudo-transition points, the critical temperature $ \beta_c $ can be determined via infinite-volume extrapolation, where we fixed $ \nu $ to its exactly known value. After averaging the individual $ \beta_c $, we arrive at
	\begin{align}
		\beta_c = 0.440688\pm 0.000015 \quad (L_\mathrm{min}=32),
	\end{align}
	where the reported uncertainty is the standard error of the average. This value is quite close to the exact critical temperature of $ \beta_{c} \approx 0.4406868$. The smallest lattice ($L=16$) is again discarded in all fits.

	The exponent $ \gamma/\nu $ is obtained from relation~\eqref{eq:scaling_chi}. Here, we exclude all lattice sizes $L \leq 64$, since residual plots indicate (slight) systematic deviations up to that value. The weighted average of the three resulting fits with acceptable quality ($Q \geq 0.3$) yields	
	\begin{align}
		\gamma/\nu = 1.7516 \pm 0.0008 \quad (L_\mathrm{min}=80)
	\end{align}
	as the final result, which is compatible with the exact value of $ 7/4 $. The residuals of all seven fits are shown in the left panel of Fig~\ref{fig:residuals-regular-gamma-and-beta}.
	
	The combinations $ (1-\beta)/\nu $ and $ \beta/\nu $ are determined from fits to the relations \eqref{eq:scaling_dmdb} and \eqref{eq:scaling_m}, respectively. For the former exponent, we find three fits with $ Q\geq 0.2 $, similar as for $ \gamma/\nu $, but the residual plots show no need to discard further data points. Our final value is thus given by the average
	\begin{align}
		(1-\beta)/\nu = 0.8747 \pm 0.0010  \quad (L_\mathrm{min}=64),
	\end{align}
	also compatible with the exact value of $ 7/8 $. However, for $ \beta/\nu $, our data does not return a single acceptable fit -- even when discarding half of the data points. A thorough analysis of the fit residuals shows no systematic corrections for $ L>32 $, but reveals that the poor quality of the fits arises from the small uncertainties assigned to the values of $ \langle m\rangle $. Indeed, the relative uncertainties are about half an order of magnitude smaller compared to, e.g., the last five observables of \eqref{eq:scaling_relations}. If we increase the uncertainties of the data points by an \emph{ad hoc} factor $ f=5 $, then five out of seven fits turn out to be acceptable, with $ Q\geq 0.01 $, producing the reasonable final average of
	\begin{align}
		\beta/\nu = 0.1261 \pm 0.0013 \quad (L_\mathrm{min}=48). 
		\label{reg_beta_nu}
	\end{align}
	
	\begin{table}[t]
		\small
		\centering
		
		\begin{tabularx}{\columnwidth}{YYX} 
			\hline
			\hline
			$\beta/\nu$ & goodness-of-fit $Q$ & at max of \\
			\hline		
			$0.1298 \pm 0.0014$ & $0.976$ & $\chi$ \\
			$0.1201 \pm 0.0014$ & $0.000$ & $C$ \\
			$0.1201 \pm 0.0013$ & $0.986$ & $\d m/\d\beta$ \\
			$0.1269 \pm 0.0017$ & $0.051$ & $\d\ln m/\d\beta$ \\
			$0.1266 \pm 0.0020$ & $0.012$ & $\d\ln m^2/\d\beta$ \\
			$0.1290 \pm 0.0015$ & $0.071$ & $\d U_2/\d\beta$ \\
			$0.1343 \pm 0.0021$ & $0.000$ & $\d U_4/\d\beta$ \\
			\hline
			$0.1261 \pm 0.0013$ &avg. $Q \geq 0.01$&\\
			\hline
			\hline
		\end{tabularx} 
		\caption{The seven single fits for $\beta/\nu$ for the regular lattice using $L = 48$ to $256$ (8 data points) as well as the average, obtained from the five values with $Q \geq 0.01$. The corresponding residuals are shown in the right panel of Fig.~\ref{fig:residuals-regular-gamma-and-beta}.}
		\label{tab:beta_full_regular}
	\end{table}

    The full list of fits can be seen in Table~\ref{tab:beta_full_regular} and the corresponding residual plot in the right panel of Fig~\ref{fig:residuals-regular-gamma-and-beta}.
    By calculating $\beta/\nu$ estimates for multiplication factors $ f=2 $ to $8 $, we observe that the number of good fits increases with $f$, but the average $\beta/\nu$ fluctuates only in the last digit, consistently maintaining compatibility with the exact result 1/8.
	
	We note that, regarding the fits for the three ratios $ \gamma/\nu $, $ (1-\beta)/\nu $, and $ \beta/\nu $, those fits that depend on the function values at the pseudo-critical points of either $ C $, $ {\rm d}\ln \langle m^2 \rangle /{\rm d}\beta $, or $ {\rm d}U_4 /{\rm d}\beta $ always present the lowest fit quality (i.e., large $ \chi_\mathrm{red}^2 $/d.o.f.). This is due to the fact that those three quantities have their maxima at a larger distance from the simulation temperature, compared to the remaining observables. Therefore, in order to obtain a larger number of acceptable fits and more accurate estimates for the critical exponents, multi-histogram reweighting methods \cite{ferrenberg1989} would be necessary. However, in the case of random lattices, the  fluctuations among replicas already prevent estimates from reaching a precision comparable to that of regular lattices, rendering more accurate reweighting methods unnecessary.

\subsection{Voronoi-Delaunay Triangulation}
\label{sect:vd}
	
	As outlined in Section~\ref{sec:VD_lattice}, a prominent example of a random lattice is given by the Voronoi-Delaunay triangulation of a Poissonian point cloud.
	Due to the spatial randomness, stronger corrections to scaling, compared to the regular case, can be expected. As a consequence, it is necessary to simulate the model on large lattices.
	For $L=16$ to $320$, we perform quenched replica averages (see Sec.~\ref{sec:Model_and_Methods}) over $N_r=1000$ independent realizations of the VD construction. For the largest lattice considered, $L=400$, only $N_r=500$ realizations are simulated. Starting from a completely ordered configuration, we perform $10^6$ cluster updates to equilibrate the system, followed by $5\cdot 10^7$ cluster updates, with measurements taken every 25th cluster update. Physical observables are obtained by reweighting for each simulated replica -- this amounts to one curve for each observable and each replica. After averaging the curves of the observables of all replicas, extremal points are determined using an iterative bisection method.	
	
	The statistical uncertainties of the replica-averaged observables are obtained from the standard error of the $N_r$ different observable curves used to calculate the average. As pointed out in Sec.~\ref{sec:Model_and_Methods}, this error estimate contains both the uncertainty corresponding to the thermal fluctuations in each replica, as well as the fluctuations among different replicas, arising from the different disorder realizations. 
	We perform linear fits to the scaling equations, as in the previous subsection, thereby ignoring any corrections to scaling. For each observable listed in Eqs.~\ref{eq:scaling_m} -- \ref{eq:scaling_dU4db}, we perform seven linear fits, each using a different estimate of the pseudo-critical temperature, as obtained from extremal points of the observables. 

	\begin{figure}[t]
		\centering
		\includegraphics[width=\linewidth]{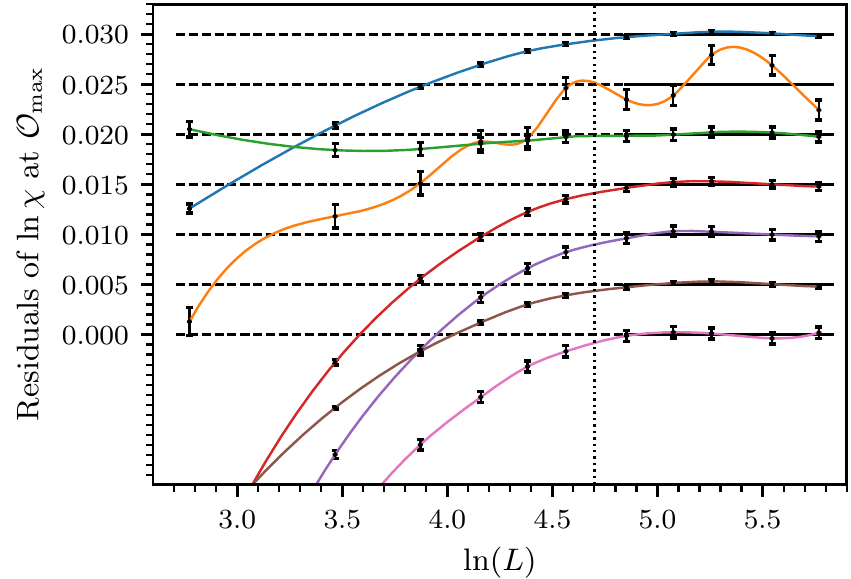}
		
		\caption{Linear fit residuals for $\gamma/\nu$ for the Voronoi-Delaunay lattice in the range from $L_\mathrm{min}=160$ to $L_\mathrm{max}=400$. The single lines in each panel correspond to observables \eqref{eq:obs_chi} to	\eqref{eq:obs_dU4db} from top to bottom and are shifted for convenience. The region of the dashed black lines show the excluded region. The curves are guides to the eye. Only the green curve shows no systematic deviation and yields $\gamma/\nu = 1.7512(7)$, with a goodness-of-fit value $Q=0.90$.}
		\label{fig:residuals-vd-gamma}		
	\end{figure}

\begin{figure}[t!]
		\centering
		\includegraphics[width=\columnwidth]{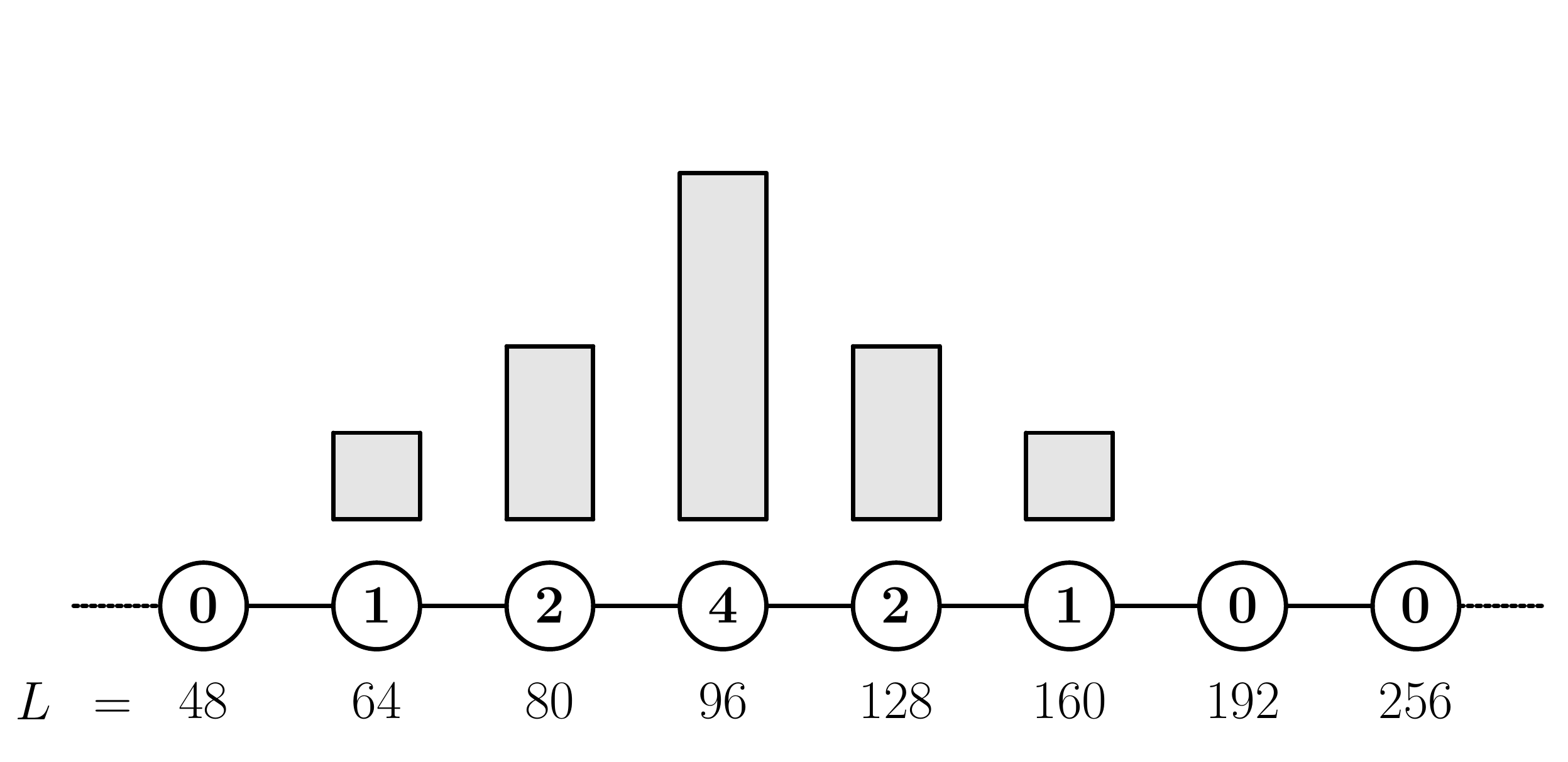}
		\caption{Schematic representation of the local fitting procedure, here for $L=96$. Each circle symbolizes one lattice size $L$, whereas the numbers inside the circle are the corresponding weights.} 
		\label{fig:weights}		
	\end{figure}

	Instead of adopting a fixed $L_\text{max}$, as for the regular lattice, we employ a local fitting procedure in order to obtain an effective exponent. More specifically, we perform the fitting over a window of five consecutive data points from the range $L \in \{ 16, 32, \dots, 320, 400 \}$, assigning weights that emphasize the central data point (see Fig.~\ref{fig:weights}). The local fitting is necessary due to the rather strong systematic deviations from a pure power-law. The residuals of the fits for $\gamma/\nu$, for instance, shown in Figure~\ref{fig:residuals-vd-gamma}, clearly demonstrate that the data points systematically deviate from the horizontal. 
	
\begin{figure*}[t]
		\centering
		\includegraphics[width=\linewidth]{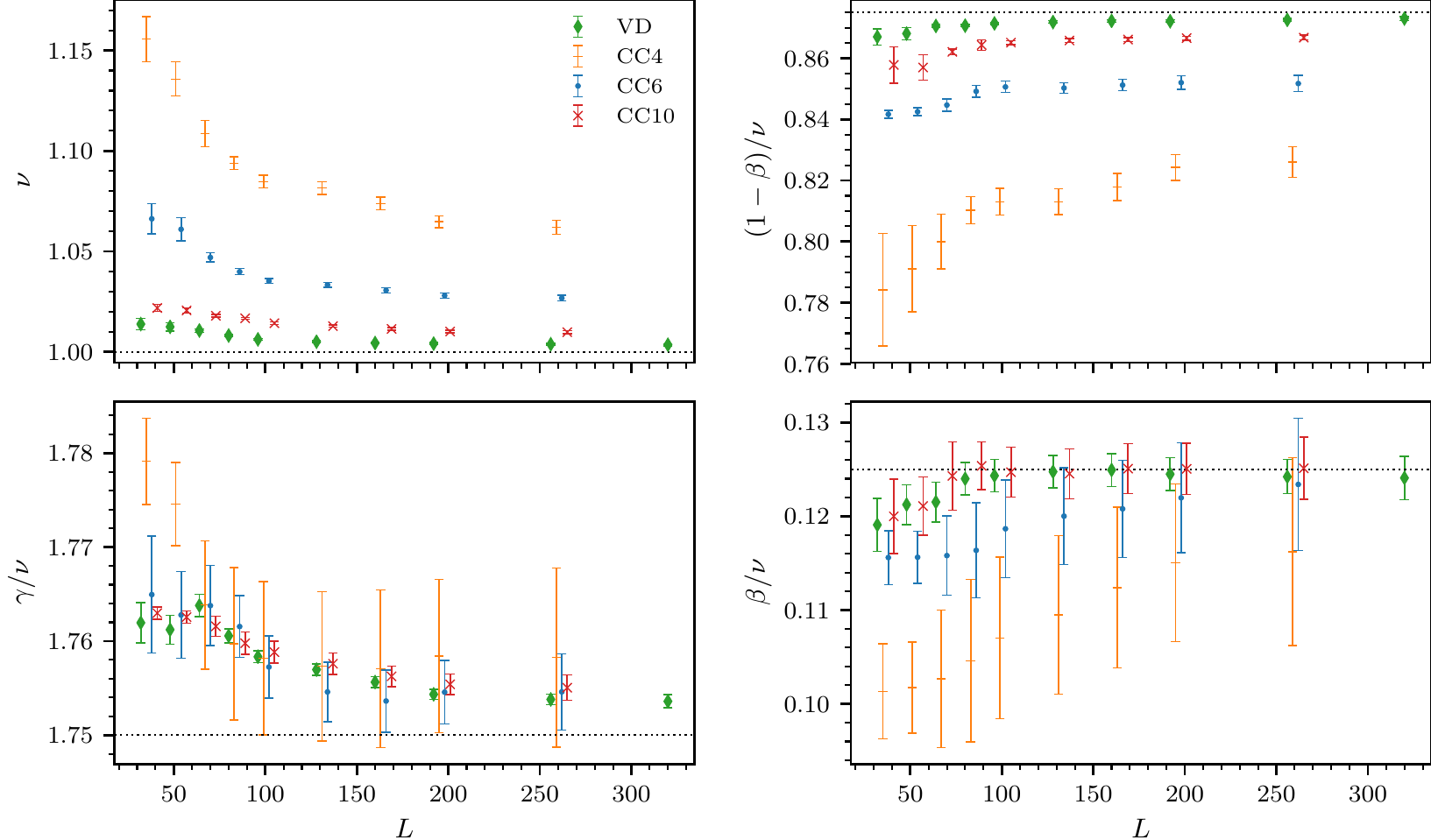}
		\caption{Comparison of critical exponent estimates for the  CC lattices with $q=4,6,10$ and VD lattices, as a function of $L$. Dashed lines indicate the clean universal values of the 2D Ising universality class.} 
		\label{fig:matrixplot}		
	\end{figure*}

	The effective exponents $\nu$, $\gamma/\nu$, $(1-\beta)/\nu$ and $\beta/\nu$ for VD are shown in Fig.~\ref{fig:matrixplot} and listed in Table~\ref{tab:fullresults}, where we display the averages of the single fits in each individual fitting window. For the estimates of $\nu$, we observe a very smooth curve, decreasing continuously as the fitting window is moved towards larger lattices. Therefore, we offer no final result for the exponent $\nu$. Regarding the estimates at hand, we expect the effective exponent $\nu$ to tend to the exact value in the infinite-volume limit. For $\gamma/\nu$, the situation is very similar. As in the case for $\nu$, the individual estimates again exhibit a systematic downwards trend and we expect the exact value to be reached in the infinite-volume limit. For the critical exponent $\beta$, which can be estimated from the scaling of $[\la m \ra ]_\mathrm{avg}$ and $\mathrm{d} [ \la m \ra ]_\text{avg} / \mathrm{d}\beta$, the corresponding curves are also smooth and indicate a tendency towards the expected values in both cases. In particular, for $ \beta/\nu$ the universal value of 0.125 is already reached within the error bars for smaller values of $L$.

\subsubsection*{Critical Temperature}	
	
\begin{figure}[t]
		\centering
		\includegraphics[width=\columnwidth]{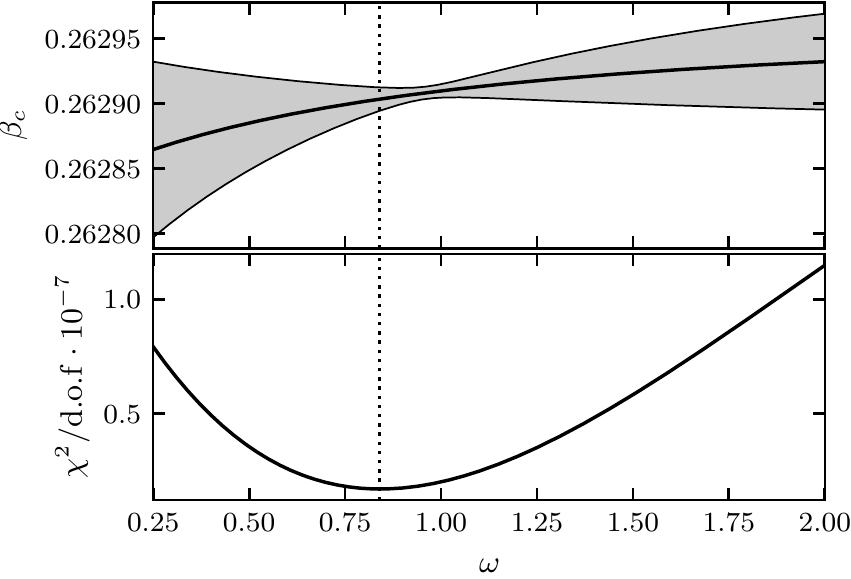}
		\caption{Top: estimate of $\beta_c$, and its error (shaded region) as a function of fixed $\omega$. Bottom: corresponding $\chi^2$ values, as described in the text. Employing different fitting algorithms generates qualitatively similar results, which are also insensitive to the choice of lattice size range.} 
		\label{fig:Tc_fit_VD}		
	\end{figure}

	Linear fits for the determination of the critical temperature according to Eq.~\ref{eq:TcScaling} reveal systematic deviations, even if many of the small lattice sizes are discarded, qualitatively similar to those observed for the exponents (compare Fig.~\ref{fig:residuals-vd-gamma}). Therefore, we decided to take into account higher order corrections in the finite-size scaling analysis in order to allow for a more precise estimate of $\beta_c$. Considering a first-order correction term, Eq.~\ref{eq:TcScaling} reads
	\begin{align}
		\label{eq:TcNonlinear}
		\beta_{\mathrm{max}_i} = \beta_c + a_i L^{-1/\nu} + b_iL^{-\omega-1/\nu},
	\end{align}
	where the correction-to-scaling exponent $\omega$ is expected to assume the ``trivial'' integer value 1,  or smaller fractional values~\cite{kaupuzs2016}. The leading correction term for the related $\phi^4$ model could have an exponent as small as $1/4$, but with amplitude too small for its effect to be measurable in moderately sized lattices. For the strongly site-diluted Ising model, which is perhaps more directly comparable to the VD model, a value of $\omega=0.63(20)$ has been found~\cite{tomita2001}. When fitting Eq.~\ref{eq:TcNonlinear} to our data, we can, in light of the results of Table~\ref{tab:fullresults}, set $\nu=1$, which reduces the number of fitting parameters to four. As the effects we are trying to detect are rather small, it is still challenging to obtain stable fits. For this reason, we perform a series of fits for different, fixed values of $\omega$ and hence obtain corresponding $\beta_c$ estimates. We follow this procedure for the data for each of the seven observables, and then calculate the average as well as the standard deviation of $\beta_c$ for each $\omega$. 
	In Fig.~\ref{fig:Tc_fit_VD}, the estimate of $\beta_c$ and its error (shaded region) is depicted together with the average reduced $\chi^2$ as a function of fixed $\omega$. It can be seen that the best fits are obtained for $\omega\lesssim1$, coinciding with the most precise estimates of $\beta_c$ as well. The best fit value is
	\begin{align}
	\beta_c=0.262904(9),
	\end{align}
	corresponding to $T_c=3.80368(13)$, at $\omega\approx0.84$. To the best of our knowledge, this is so far the most precise value available for the critical coupling for the 2D Ising model on a Voronoi-Delaunay lattice.
	
\subsection{Constant Coordination Model}
\label{sect:CCresults}

	We study CC lattices for $q=4,6$ and $10$ (short: CC4, CC6, CC10) on a Poissonian point process. For CC4, CC6 and CC10, we use the same number of independent disorder realizations, measurements, equilibration steps and cluster updates as for the VD lattice. In this way, the results for the different models are of comparable precision. The estimates of the exponents are obtained following the same procedures described in Sec.~\ref{sect:vd}.
	The results for the CC4, CC6 and CC10 models are presented in Fig.~\ref{fig:matrixplot}, where we also added the VD exponents for comparison. A detailed list of the data points can also be found in Table~\ref{tab:fullresults}.
			
	Recall that due to the nature of the CC construction small isolated components may occur, in contrast to the VD construction, where the lattice always consists of one single component. In order to properly update those islands, we employ an additional Metropolis step between measurements. Furthermore, we report results only for lattices of size $L=16$ to $L=320$, as for larger lattice sizes the construction already becomes as expensive as the actual Monte Carlo simulation.  	
	
	Overall, in Fig.~\ref{fig:matrixplot} we see similar tendencies as for the VD lattices, however, with larger absolute deviations from the universal Ising values. For the correlation length exponent $\nu$, all CC models seem to show a systematic trend. However, the deviations become larger for smaller $q$. Compared to VD, for $q=10$ those deviations are roughly three times as large, and for $q=6$ already about one order of magnitude higher. Considering also the fact that for $q=6$ the error bars are only about twice as large as for VD, the results indicate that any possible convergence is significantly slower.
	
	A remarkably different situation arises for the susceptibility exponent $\gamma/\nu$, where the effective exponents for all models seem to collapse. However, whereas the CC10 shows a clear trend of decreasing estimates, this behavior becomes less distinct for $q=6$, where the curves seem to saturate within the considered range of  $L$. Eventually, for $q=4$, almost all values are compatible with 7/4, which may, however, be a consequence of the relatively large error bars.
	
	As for $\nu$, the exponent $(1-\beta)/\nu$ is clearly different for CC and VD graphs. For the VD construction, a trend consistent with $(1-\beta)/\nu=0.875$ is evident. In contrast, the CC exponents are further away from the universal value and show no clear trend, with a possible exception of CC4, where the effective exponent appears to increase with $L$. Similar to $\nu$, the absolute deviations for CC6 are already almost one order of magnitude larger than for VD.	
	Finally, the exponent $\beta/\nu$ shows a clear trend towards the universal value in the case of $q=4$ and $6$, with the last few data points being fully compatible with 1/8. For $q=10$, however, all estimates match the universal value, very similar to the VD model.

	For completeness, we state the critical temperatures for the constant coordination models which are roughly $\beta_c\approx0.549$, $0.294$, and $0.148$ for CC4, CC6, and CC10, respectively. A more precise determination of the model $T_c$, if necessary, could be obtained using the methods described in Sec.~\ref{sect:vd}, but is omitted, since these values depend on the fine-tuning of the CC algorithm and are therefore non-universal.
	
	We also consider the CC model on the hard core point sampling (CC6-HCPP), where only 320 disorder realizations have been used. We found that more ordered lattice results in critical exponents closer to universal ones, as can be seen in Appendix~\ref{sect:hcpp}.
\section{Discussion}	
\label{sec:Discussion}

	In comparison with the VD triangulation, the CC model for $q=6$ has exponents $\nu$ and $(1-\beta)/\nu$ that show deviations from their respective universal values which are larger by about one order of magnitude (e.g, $\nu=1.0042(5)$ for VD and $\nu=1.0281(12)$ for CC6 at $L=192$). Furthermore, for all CC models, the convergence of the effective exponents seems weaker or even doubtful, with the possible exception of $\beta/\nu$, which has rather large relative errors. 
	
	In the following, we want to understand our findings using a number of topological arguments. We start by referring again to \cite{barghathi2014}, where it was shown that for the VD lattice, the constrained total coordination number imposes strong anti-correlations in the local $q$ fluctuations, which in turn are responsible for the fast decay of disorder under spatial renormalization-group-type blocking transformations (compared to, e.g., diluted lattices) and are thus asymptotically irrelevant. It was reasoned that this fast decay can be expressed in terms of a modified Harris criterion $(d+1)\nu>2$ that explains the fact that, e.g., simulations of the contact process on those lattices show the clean universal exponents \cite{oliveira2008}, although the classical Harris criterion $d\nu>2$ is violated.
	
	In the present paper, we construct a lattice which provides random connectivity (and thus topological disorder)\footnote{Keep in mind that \emph{random connectivity} does not imply a \emph{random coordination number} at this point.} and -- as an obvious effect of the fixed local coordination number -- no fluctuations in the original lattice or on any blocking level. Therefore, since the effective critical exponents clearly deviate from the corresponding universal values, we are led to the conclusion that the scaling of $\sigma_Q$  under coarse-graining should not be the decisive property determining the nature of the phase transition. This conclusion is supported by the very recent results of~\cite{schawe2017}, where it is shown that the universality of the 2D Ising model on GG and RNG is unchanged and therefore belongs to the same class as the Ising model on a regular lattice. In Sec.~\ref{sec:LatticeModels}, we perform the blocking analysis using Eq.~\ref{eq:flucScale1} for these two types of proximity graphs and find that both of them unambiguously show  a decay of $\sigma \propto L_b^{-1}$. This means that disorder in these graphs decreases as slow as for generically disordered models. Hence, the results from \cite{schawe2017} are not covered by the modified Harris criterion.
	
	{\setlength{\extrarowheight}{2pt}
		\begin{table*}[htp]
			\small
			\centering
			\begin{tabular}{C{13.4mm} C{23mm} C{32mm} C{32mm} C{13mm} C{15mm} C{22mm} L{17mm}}
				\hline
				\hline
				Model  &   Disorder decay\newline exponent & Coordination number \newline anti-correlation & Total coordination \newline number constrained & Planar & Connected & 2D Ising \newline universality & \\ 
				\hline
				VD      & $3/2$     & yes$^{\whitedagger}$ & yes  & yes    & yes       & yes& \cite{barghathi2014}  \\ 
				CC      & \textemdash                   & no$^{\whitedagger}$                      & yes     & no     & no        & questionable &\\ 
				GG      & $1^\dagger$         & no$^\ddagger$  & no      & yes    & yes       & yes &\cite{schawe2017} \\ 
				RNG     & $1^\dagger$        & yes$^\ddagger$ & no      & yes    & yes       & yes &\cite{schawe2017} \\ 
				BD      & $1^{\whitedagger}$       & no$^{\whitedagger}$                     & yes     & yes    & no        & strong/weak& \cite{barghathi2014}, \cite{dotsenko1981,jug1983,shankar1987,shalaev1994,ludwig1987,kim1994a,kim1994b,selke1994b,ziegler1994,kuehn1994,kim2000,gordillo2009,martins2007,fytas2010,fytas2013,zhu2015} \\
				SD      & $1^\dagger$         & yes$^{\whitedagger}$            & no            & yes     & no       & strong/weak&\cite{dotsenko1981,jug1983,shankar1987,shalaev1994,ludwig1987,kim1994a,kim1994b,selke1994b,ziegler1994,kuehn1994,kim2000,gordillo2009,martins2007,fytas2010,fytas2013,zhu2015} \\
				RGG     & $1^\dagger$         & no$^{\whitedagger}$             & no             & no     & no        & unclear$^\S$ &\\ 
				qNNsym & $1^\dagger$ & no$^\ast$ & no & no & no & unclear$^\S$ & \\
				\hline
				\hline
			\end{tabular} 
			\caption{Overview of the lattice models considered in this work. The disorder decay exponent $a$ is defined by the relation $\sigma_Q \sim L_b^{-a}$, where $\sigma_Q$ is defined according to Eq.~\ref{eq:flucScale1}. The values marked with $\dagger$ have been calculated in Sec.~\ref{sec:LatticeModels}, see Fig.~\ref{fig:blocking}, the symbol $\ddagger$ refers to Fig.~\ref{fig:Corr} (upper panel). $\S$: for the RGG and the symmetric qNN, no clear universal properties are expected, see text. {\footnotesize $\ast$}: negative correlations are present, but positive correlations are dominant, especially on the typical scale of one bond length, see Fig.~\ref{fig:Corr}.}
			\label{tab:overview}
		\end{table*}
	}	
	
	We collect several types of disordered lattice models in Table~\ref{tab:overview}, together with some relevant geometric properties and statements concerning the universality of the 2D Ising model on each lattice. From the overview given in this table, we claim that the general statement of topological disorder being less relevant than generic disorder, as stated in Ref~\cite{barghathi2014}, is perhaps too general. However, the particular instances of lattices mentioned by the authors can indeed be expected to preserve the universal features of a transition, since they are all tilings. The key difference between tilings and lattices with bonds that may cross each other (like our CC model or the random geometric graph with fixed interaction radius) lies in the fact that for tilings, it is always ensured that there exists one single component containing all sites. We thus conclude that very clear universal properties (e.g., no strong logarithmic corrections) are obtained if the underlying lattice is both \emph{planar and connected}\footnote{Tilings are a special case of planar, connected graphs.}. Here, we remind that a graph is called planar if it \emph{can be} embedded in the plane such that there are no edge crossings. Whether a specific graph is planar can be checked according to Kuratowski's theorem \cite{kuratowski1930}. Since  RNG and GG possess these properties, this would explain the positive results from \cite{schawe2017}.
	
	Comparing the GG with RGG (see Table~\ref{tab:overview}), it is clear that -- apart from the RGG being neither planar nor connected -- they show the same geometric characteristics. Following our line of argumentation, the Ising model on the RGG lattice is expected to have disorder dependent effective critical exponents, exactly as for the CC model (see Sec.~\ref{sect:CCresults}). Some preliminary simulations with a finite interaction radius of $R_{\bar{q}=6}$ (see Eq.~\ref{eq:CaptureRadius}), not presented here, indeed seem to confirm this expectation. Similar holds for the symmetrized $q$-Nearest-Neighbor graph, see also Tab.~\ref{tab:overview}.
	
	Moreover, a prominent and well-studied example of disordered lattices that are planar but not connected are the site- or bond-diluted regular lattices (see Fig.~\ref{fig:geometry}), also included in Table~\ref{tab:overview}. They allow for isolated clusters and thus show a percolation transition, resulting in a multicritical point in the temperature/dilution-probability phase diagram. The constant coordination model also allows for the occurrence of isolated islands. By employing a cluster counting procedure, we calculate the fraction $1-p_\mathrm{con}$ of all sites on the CC lattices belonging to islands disconnected from the giant component. For the CC4, we find $1-p_\mathrm{con}\sim 10^{-3}$. As expected, this number decreases strongly as $q$ is increased. For $q=6$ we find $1-p_\mathrm{con}\sim 10^{-6}$ and for $q=10$ no small island were detectable in all of the $N_r=1000$ realizations of constant coordination lattices with $L=320$, yielding $1-p_\mathrm{con}< 10^{-7}$ as an upper bound\footnote{Note that, due to the constraint of fixed $q$, the smallest possible isolated component needs to contain at least 11 sites. Thus, if we had found one single of them in the 1000 realizations the fraction would have been calculated by $11/(1000\cdot320^2)\approx 10^{-7}$.}. Considering, in contrast, smaller values of $q$, say $q=2$, the lattice would undergo a percolation transition, as in this case the formation of, e.g., triangles (3 sites, 3 links) is very likely and a giant component may not form at all in most realizations.
	
	It should be emphasized that the effect of the isolated islands on the measured observables might be negligible, since, even for CC4 lattices, such sites amount to only to $0.1\%$ of the total lattice sites. Furthermore, we ensured that isolated clusters are properly updated by local Metropolis updates, as explained in Sec.~\ref{sect:singSquare}. By decreasing $q$ below the percolation threshold, though, any collective long-range magnetic phase must inevitably be destroyed since the system is then decomposed into many disconnected finite clusters and no collective long-range behavior can be maintained. 
	
	Reviewing the ample literature on the 2D site- or bond-diluted Ising model, one indeed finds remarkable similarities to our results for CC. First of all, many numerical simulations seem to show exponents which are clearly non-universal and vary dramatically with dilution strength. Already in the 1990's, these numerical results, as well as field-theoretic calculations, led to a controversy that still persists, regarding the universal character of those models. According to the so-called \emph{strong universality} hypothesis, disorder is marginally irrelevant, leading to clean exponents accompanied by logarithmic scaling corrections and, particularly remarkable, a specific heat diverging ultra-slowly in form of a double logarithm~\cite{dotsenko1981,jug1983,shankar1987,shalaev1994,ludwig1987}. The \emph{weak universality} scenario, in contrast, posits leading critical exponents that vary continuously with the strength of the dilution, but with some quotients of exponents, such as $\gamma/\nu$ and $\beta/\nu$, remaining unchanged \cite{kim1994a,kim1994b,selke1994b,ziegler1994,kuehn1994,kim2000}. For a comprehensive historical review covering articles supporting either of the two scenarios, we refer the reader to \cite{gordillo2009}. Currently, the strong universality scenario is favored, having been strengthened by recent numerical studies \cite{martins2007,fytas2010,fytas2013}, with Zhu et al.~effectively ruling out the weak scenario for their large-scale, high-accuracy results \cite{zhu2015}.
		
	Comparing our results with those from the aforementioned studies of diluted models, we recognize a number of similarities. In particular, the effective exponents $\nu$ and $(1-\beta)/\nu$ change continuously with the lattice parameter $q$, whereas $\gamma/\nu$ varies only slightly among the models and $\beta/\nu$ is already compatible with the universal value for all choices of $q$. Given these similarities, the question arises whether topological disorder in the CC model is also marginally irrelevant and logarithmic corrections arise (i.e., strong universality) or whether one is facing continuously varying leading critical exponents (as proposed by the weak universality hypothesis). As both scenarios predict unchanged values for the ratios $\gamma/\nu$ and $\beta/\nu$, they both can not be used for a distinction. The specific heat, in contrast, shows a different scaling behavior already in the leading order. For the strong scenario, a double-logarithmic scaling 
	\begin{align}
	C = a\ln(b\ln(cL))
	\label{eq:strong_universality}
	\end{align}
	can be expected \cite{hasenbusch2008a,mazzeo1999,kenna2008}, whereas weak universality would predict a power-law scaling 
	\begin{align}
	C = C_0 + aL^{\alpha/\nu}
	\label{eq:weak_universality}
	\end{align}
	with negative exponent. In order to investigate the origin of the deviations from clean universality in our models, we fit the finite-size data of the specific heat $C$ to Eqs.~\ref{eq:strong_universality} and \ref{eq:weak_universality}, summing up to seven fits each (corresponding to the maxima of the observables \eqref{eq:scaling_relations}). Remarkably, both scenarios fit the data equally well. Even when including the smallest lattice size, $L=16$, we find reduced $\chi^2$ values between 0.5 and 3 for all seven fits for both fitting functions. However, as the leading scaling behavior is only valid for large $L$, we discard the smallest lattice size which again significantly increases the quality of most fits. Specifically, for the strong scenario, Eq.~\ref{eq:strong_universality}, we then find 5 out of 7 fits with reduced $\chi^2$ in the range 0.1 to 0.2. Furthermore, if further lattice points are discarded, all fits appear very stable. For the power-law scenario, Eq.~\ref{eq:weak_universality}, after discarding $L=16$, we also find 5 out of 7 fits with very good quality (see Tab.~\ref{tab:alpha_cc}). Moreover, the fits are again numerically very stable and their quality (in terms of $\chi^2$) as well as the the fitted parameters $\alpha/\nu, a$ and $ C_0$ show no systematic trend if further lattices are discarded. As can be seen in Tab.~\ref{tab:alpha_cc}, all seven fits consistently yield a small negative exponent $\alpha/\nu$. Performing a simple average with standard error, we obtain an exponent ratio of $ \alpha/\nu = -0.048(12) $ for the CC6 model. Using the hyperscaling relation $2-\alpha=d\nu$, this yields a correlation length exponent of $ \nu=1.025(6) $, which is, rather remarkably, compatible with the effective exponent $\nu$ we obtain in Sec.~\ref{sect:CCresults} for the largest lattices available (see also top left panel of Fig.~\ref{fig:matrixplot}). In light of these findings one may speculate about whether the 2D Ising model on the CC lattice is situated in a weakly universal scenario with $q$-dependent leading exponents. The exponents $ \nu $ and $(1-\beta)/\nu$ in Fig.~\ref{fig:matrixplot} would thus not tend towards the respective clean universal value. However, it should be emphasized again that also the logarithmic corrections fit the data well. Therefore, we can not ultimately decide on either scenario.
	
	
	\begin{table}[t]
		\small
		\centering
		
		\begin{tabularx}{\columnwidth}{YYYYX} 
			\hline
			\hline
			$\alpha/\nu$ & $ C_0$ & $a$ & red. $ \chi^2 $ & at max of \\
			\hline		
			$ -0.026 $ & 13.58 & $-13.21$ & 0.11 & $\chi$\\
			$ -0.040 $ & 9.78  & $-9.42$  & 1.95 & $C$\\
			$ -0.052 $ & 8.00  & $-7.70$  & 1.20 & $\d m/\d\beta$\\
			$ -0.059 $ & 7.20  & $-7.14$  & 0.07 & $\d\ln m/\d\beta$\\
			$ -0.066 $ & 6.61  & $-6.66$  & 0.10 & $\d\ln m^2/\d\beta$\\
			$ -0.043 $ & 9.06  & $-8.84$  & 0.08 & $\d U_2/\d\beta$\\
			$ -0.050 $ & 8.07  & $-8.04$  & 0.21 & $\d U_4/\d\beta$\\
			\hline			
			\hline
		\end{tabularx} 
		\caption{Single fits for the specific heat finite-size data to Eq.~\ref{eq:weak_universality} for the CC6 lattice with $L$ ranging from $32$ to $320$ (10 data points). }
		\label{tab:alpha_cc}
	\end{table}

	Interestingly, the Ising model is clearly consistent with the universal critical exponents when placed on a CC lattice built from a HCPP (cf.~Appendix~\ref{sect:hcpp}), instead of a fully random distribution. The moderate ordering, arising from the repulsive character of the hard sphere model, has two major effects. First, the probability that a small group of sites does not belong to the giant component can be considered virtually zero. Second, due to the more homogeneous distribution, the number of bonds crossing each other is significantly reduced. Hence, the lattice becomes increasingly more similar to a tiling as the degree of repulsion is increased.

\section{Conclusion}
\label{sec:Conclusion}

	We study the two-dimensional Ising model under a novel type of topological disorder, namely a random lattice whose local coordination number $q$ is constant throughout the system. This construction allows us to eliminate the influence of coordination number fluctuations on the phase transition, which in previous studies has been referred to as \emph{the} relevant quantity determining whether the universal properties are preserved. By keeping $q$ locally (and therefore also globally) fixed, we are thus able to study disorder from a different perspective.
	In particular, we propose a dynamical method to construct constant coordination lattices, where pairs of bonds are minimized with respect to their lengths until the desired degree of locality is reached. Disorder is therefore solely encoded in the neighbor relations among the points. On three particular types of those lattices, we conducted large-scale Monte Carlo simulations of the Ising model and determined effective critical exponents with high accuracy using finite-size scaling relations. The calculations are compared to simulations of the Ising model on Voronoi-Delaunay lattices. 
	
	In summary, although the coordination number is fixed, we observe rather large fluctuations in the individual transition temperatures among the independent disorder realizations. Furthermore, similarly to generically disordered lattices (e.g., diluted systems), some of the critical exponents seem to vary with the disorder strength. Applying a logarithmic, as well as a weakly universal scaling scenario to the specific heat, we find that both scenarios fit the data well. Therefore, the exact origin of the deviations remains undecided. This, in light of other recent results, can be seen as a strong indication that fluctuations in the coordination number do not exclusively determine the stability of the phase transition against quenched disorder.
	
	Instead, we conjecture that the lattice topology needs to be planar and connected in order to ensure clear universal properties. One natural next step would be to study a lattice which is connected, but \emph{not} planar in order to figure out whether planarity is really a necessary condition or whether connectedness is already sufficient. One such lattice would be a VD$^+$ lattice: a Delaunay triangulation with additional local random bonds.
	However, the two-dimensional Ising model is a marginal case in terms of the Harris criterion, which makes it challenging or even impossible to discriminate between universal and non-universal behavior (see corresponding Table~\ref{tab:overview}). For the diluted models, this has led to an ongoing debate about whether the exponents are truly universal or depend on the disorder strength. For this reason, we address in an upcoming publication the 2D contact process as one particular realization of the directed percolation universality class. It is known that this non-equilibrium model behaves dramatically different on diluted lattices, including an exotic infinite-randomness critical point with activated dynamical scaling as well as strong Griffith singularities (see, e.g., \cite{vojita2005,vojita2009}) compared to a clean critical behavior on the VD triangulation~\cite{oliveira2008}. The simulations of the contact process on the lattices of Table~\ref{tab:overview} should therefore allow to determine the universality character of second-order phase transitions on quenched topological disorder.

\begin{acknowledgments}
	We thank H.~Hinrichsen and T.~Vojta for helpful discussions. We also thank I.~Rausch from Ghent University for sharing his expertise on spatial random networks with us. J.S.E.P. thanks H.~Hinrichsen and the University of W\"urzburg for the hospitality and the UTFPR for supporting his research visit. M.S.~thanks the Studienstiftung des deutschen Volkes for financial support. This work is part of the DFG research project Hi~744/9-1.
\end{acknowledgments}

\setcounter{table}{0}
\setcounter{figure}{0}   
\setcounter{equation}{0}   
\renewcommand{\thetable}{\thesection.\arabic{table}}
\renewcommand{\thefigure}{\thesection.\arabic{figure}}
\renewcommand{\theequation}{\thesection.\arabic{equation}}
\appendix

	\section{Time Complexity of the CC Construction}
	\label{sect:sa}
	
\begin{figure}[!h]
		\centering
		\includegraphics[width=0.49\linewidth]{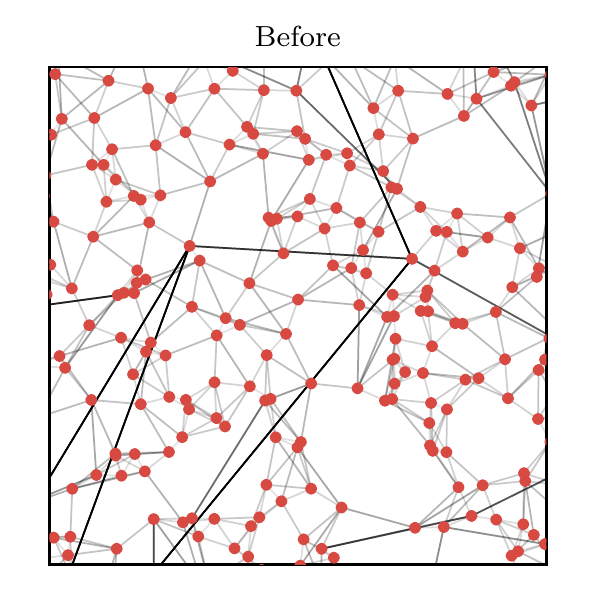}
		\includegraphics[width=0.49\linewidth]{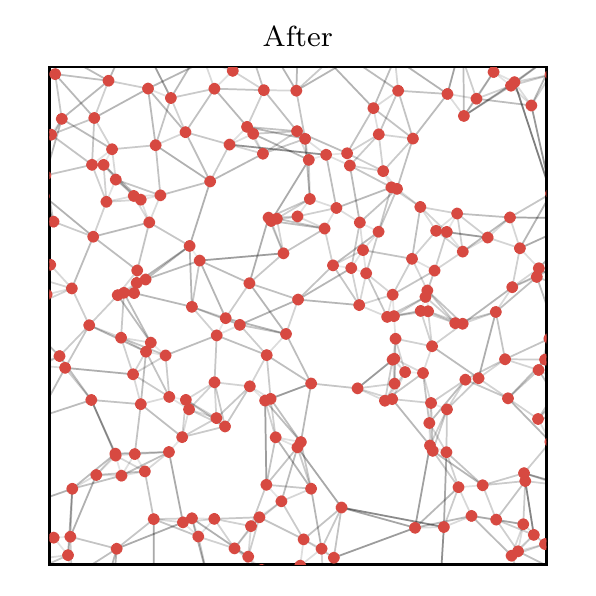}
		\caption{Lattice configurations before and after the simulated annealing process. The long bonds, highlighted in dark, have been effectively rearranged.} 
		\label{fig:appendixFIG}		
	\end{figure}

	As explained in Sec.~\ref{sec:CC}, we use a simulated annealing method for the dynamical construction of the Constant Coordination lattice. We should point out that the outlined method is rather expensive in comparison with typical $ \mathcal{O}(N \ln N) $ methods, such as for the VD triangulation and the construction of an RGG using a search tree. For the CC lattice, a scaling of $ \mathcal{O}(N^2) $ can not be avoided, since lattices of all sizes should be processed  to the same degree. Furthermore, since Eq.~\ref{eq:constraint} is a quite severe constraint, the majority of update attempts will be rejected if one applies the naive approach of a simple trial-and-error. The acceptance rate can easily be increased by optimizations, e.g., by picking the second link not completely at random, but in the local neighborhood of the link selected first. However, this changes only the constant prefactor of the asymptotic $ \mathcal{O}(N^2) $. Another optimization that we use is to start not from a fully random lattice, but from an initial lattice where the sites are connected to their nearest neighbors until the chosen $q$ of each site is reached. In Figure~\ref{fig:appendixFIG}, where a coordination number of $q=6$ was chosen, a comparison of the bond configurations before and after the simulated annealing indicates that the algorithm works as intended. All bonds have been shortened effectively. In our simulations, we performed $  \alpha\, q\, N_s\, N^{2}$ rewiring attempts for all of our lattices, where the prefactor $\alpha\approx0.01$ is found to be sufficient to achieve the desired degree of locality and $ N_s\approx 30 $ denotes the number of (logarithmic) temperature steps looped over.

\setcounter{table}{0}
\setcounter{figure}{0}   
\setcounter{equation}{0}  
	
\section{Hard Core Point Process}
\label{sect:hcpp} 
	
A more ordered variant of the CC lattice can be obtained by starting from a Hard Core Point Process (HCPP) instead of a Poisson Point Process (PPP) \cite{baddeley2015}. In the hard core model, also known as \emph{Poisson disk sampling}, the random points are placed respecting a minimum distance $R_r$ from each other. Such distributions find wide application in computer graphics, especially in sampling algorithms, and can be generated through several methods~\cite{lagae2008}, the simplest of which is \emph{dart throwing}, whereby randomly drawn points that do not satisfy the minimum distance requirement are discarded. 
There are more efficient methods, but we adopt dart throwing nonetheless, since it is easily generalized to higher dimensions~\cite{bridson2007} and the point distribution generation takes only a small fraction of the run time of our simulations.
The minimum distance $R_r$ cannot exceed the closed packing value of $(1/\sqrt[4]{12})L^{-1} \approx 0.54L^{-1}$, which corresponds to the hexagonal lattice~\cite{musin2016}. For a PPP, though, a large fraction of the construction attempts fails for $R_r\gtrsim 0.41L^{-1}$.	
In practice, for a given $N$, we set $R_r=(3/8) L^{-1}=0.375N^{-1/2}$, which results in a good coverage without being so close to the densest packing as to be overly expensive. A sample of the CC neighbor construction on this hard core point process can be seen in Fig.~\ref{fig:geometry}.

\begin{figure}
	\centering
	\includegraphics{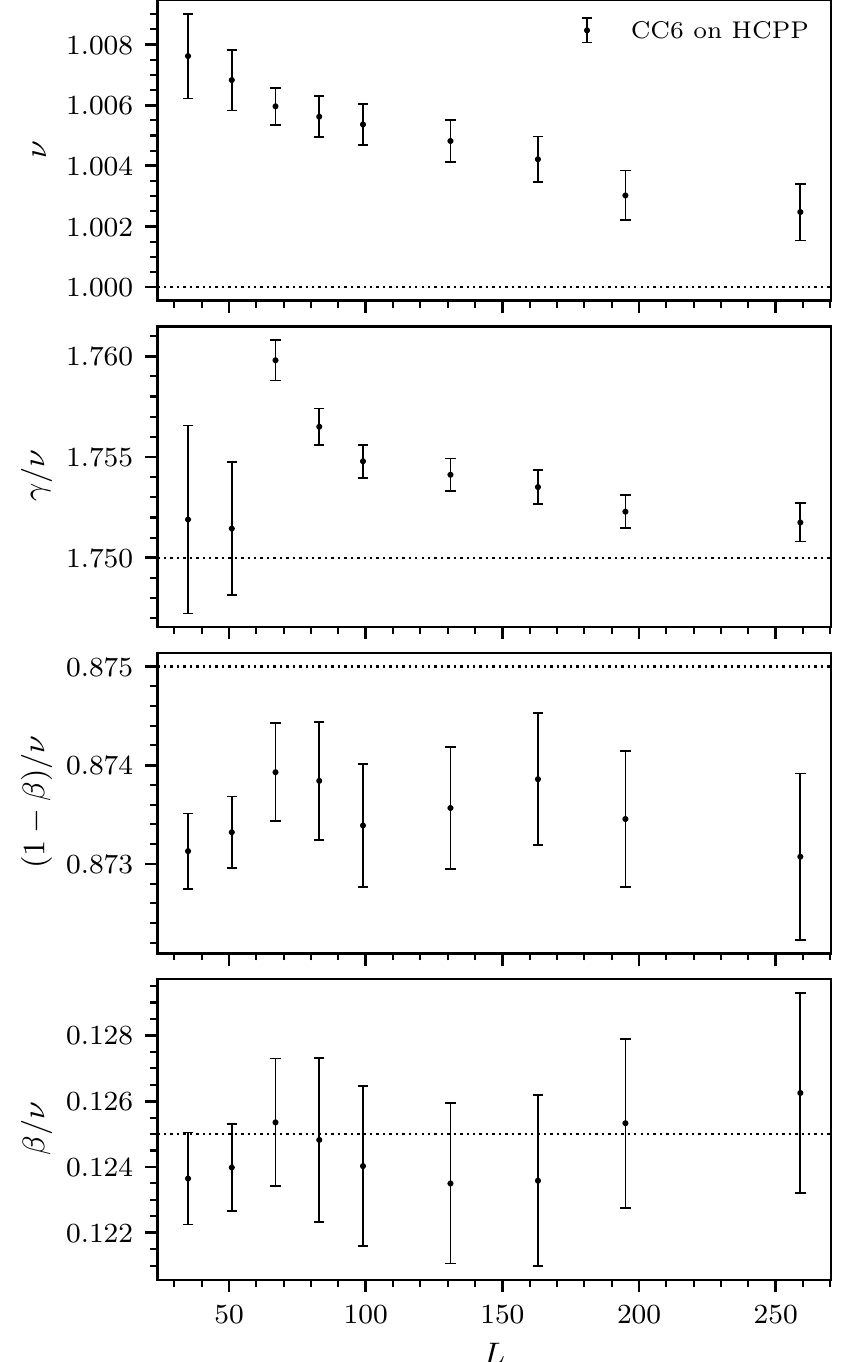}
	\caption{Effective critical exponent estimates for the CC6-HCPP model as a function of the center of the fitting window, $L$. Dashed lines indicate the clean universal values of the 2D Ising universality class.}
	\label{fig:matrixplot_small}		
\end{figure}

The effective exponents we obtained are shown in Fig.~\ref{fig:matrixplot_small} and listed in Table~\ref{tab:fullresults}. Note that the scales of the ordinate axis in the figure differ by up to one order of magnitude from the ones in Fig.~\ref{fig:matrixplot}, thus already indicating much smaller deviations from the universal values. In particular, in the top panel, where $\nu$ is depicted, we find a smooth behavior and a deviation of less than $ 0.8\%$ from $\nu=1$ even for small lattices. Moreover, the results indicate that the universal value is approached significantly faster for large $L$, compared to all other disordered models, including VD -- although a direct comparison can not be made, given that the starting distributions are statistically different. A similar result is observed for $\gamma/\nu$, where the convergence towards the universal value is also very pronounced and notably faster than in the VD case. 
Interestingly, once more, a different situation arises for $(1-\beta)/\nu$, where the values lie closer to 7/8 with only about $0.1\%$ to $0.2\%$ deviation, though they seem to remain in that range with no visible trend in either direction for larger lattices. This is in qualitative agreement with the other CC models. Finally, for $\beta/\nu$, all of the data points match the corresponding universal value within error bars. 

\setcounter{table}{0}
\setcounter{figure}{0}   
\setcounter{equation}{0}  

\section{Tables}

Table~\ref{tab:fullresults} lists the effective critical exponents of the models VD, CC4, CC6, CC10 and CC6-HCPP.

\newcommand\lenA{15mm}
\newcommand\lenB{12mm}
{\setlength{\extrarowheight}{-0.1pt}
	\begingroup
	\squeezetable
	\setlength\tabcolsep{2pt}
	\begin{table*}[!ht]
		\small
		\centering
		\begin{ruledtabular}
			\begin{tabular}{L{17mm}lC{16mm}lC{\lenB}C{\lenA}C{\lenB}C{\lenA}C{\lenB}C{\lenA}C{\lenB}C{\lenA}} 
				& $L$  & $N_\mathrm{fits}$ & $\nu$ & $N_\mathrm{fits}$ & $\gamma/\nu$ & $N_\mathrm{fits}$ & $(1-\beta)/\nu$ & $N_\mathrm{fits}$ & $\beta/\nu$ \\ 	
				\colrule
				VD &&{\footnotesize$Q>0.2$}&&{\footnotesize$Q>0.2$}&&{\footnotesize$Q>0.2$}&&{\footnotesize$Q>0.8$}& \\
				\colrule
				&	32	&	4	&	1.0139(28)	&	3	&	1.7620(23)	&	2	&	0.8670(28)	&	2	&	0.1191(28)	\\
				&	48	&	4	&	1.0125(2)	&	3	&	1.7612(16)	&	2	&	0.8681(20)	&	3	&	0.1212(22)	\\
				&	64	&	16	&	1.0107(9)	&	5	&	1.7638(12)	&	6	&	0.8706(6)	&	5	&	0.1215(21)	\\
				&	80	&	25	&	1.0082(5)	&	6	&	1.7606(7)	&	7	&	0.8707(4)	&	6	&	0.1240(17)	\\
				&	96	&	28	&	1.0063(5)	&	7	&	1.7583(6)	&	7	&	0.8714(4)	&	6	&	0.1243(17)	\\
				&	128	&	28	&	1.0051(4)	&	7	&	1.7570(6)	&	7	&	0.8719(4)	&	6	&	0.1247(17)	\\
				&	160	&	28	&	1.0045(5)	&	7	&	1.7556(6)	&	7	&	0.8722(4)	&	6	&	0.1249(17)	\\
				&	192	&	28	&	1.0042(5)	&	7	&	1.7543(5)	&	6	&	0.8721(4)	&	5	&	0.1245(17)	\\
				&	256	&	28	&	1.0038(5)	&	7	&	1.7538(6)	&	7	&	0.8726(5)	&	5	&	0.1242(18)	\\
				&	320	&	27	&	1.0036(6)	&	6	&	1.7536(7)	&	6	&	0.8730(7)	&	4	&	0.1241(23)	\\
				\colrule
				CC4 &&{\footnotesize$Q>0.01$}&&{\footnotesize$Q>0.2$}&&{\footnotesize$Q>0.3$}&&{\footnotesize$Q>0.6$}& \\
				\colrule
				&	32	&	7	&	1.156(11)	&	7	&	1.779(5)	&	3	&	0.784(19)	&	7	&	0.101(5)	\\
				&	48	&	3	&	1.136(8)	&	7	&	1.775(4)	&	3	&	0.791(14)	&	7	&	0.102(5)	\\
				&	64	&	19	&	1.109(6)	&	7	&	1.764(7)	&	5	&	0.800(9)	&	7	&	0.103(7)	\\
				&	80	&	28	&	1.0939(32)	&	7	&	1.760(8)	&	7	&	0.810(5)	&	7	&	0.105(9)	\\
				&	96	&	28	&	1.0847(31)	&	7	&	1.758(8)	&	7	&	0.813(4)	&	7	&	0.107(9)	\\
				&	128	&	28	&	1.0815(31)	&	7	&	1.757(8)	&	7	&	0.813(4)	&	7	&	0.110(8)	\\
				&	160	&	28	&	1.0739(31)	&	7	&	1.757(8)	&	7	&	0.818(5)	&	7	&	0.112(9)	\\
				&	192	&	28	&	1.0648(30)	&	7	&	1.758(8)	&	7	&	0.824(4)	&	7	&	0.115(8)	\\
				&	256	&	28	&	1.0619(35)	&	7	&	1.758(10)	&	7	&	0.826(5)	&	7	&	0.116(10)	\\
				\colrule
				CC6 &&{\footnotesize$Q>0.05$}&&{\footnotesize$Q>0.1$}&&{\footnotesize$Q>0.3$}&&{\footnotesize$Q>0.6$}& \\
				\colrule
				&	32	&	5	&	1.066(8)	&	4	&	1.765(6)	&	6	&	0.8417(13)	&	4	&	0.1156(29)	\\
				&	48	&	5	&	1.061(6)	&	4	&	1.763(5)	&	6	&	0.8425(13)	&	7	&	0.1156(28)	\\
				&	64	&	18	&	1.0470(23)	&	7	&	1.764(4)	&	6	&	0.8447(20)	&	7	&	0.119(4)	\\
				&	80	&	24	&	1.0399(16)	&	7	&	1.7615(33)	&	7	&	0.8492(19)	&	7	&	0.116(5)	\\
				&	96	&	28	&	1.0353(13)	&	7	&	1.7572(33)	&	7	&	0.8506(19)	&	7	&	0.1190(5)	\\
				&	128	&	28	&	1.0334(12)	&	7	&	1.7546(32)	&	7	&	0.8503(18)	&	7	&	0.120(5)	\\
				&	160	&	28	&	1.0307(12)	&	7	&	1.7536(33)	&	7	&	0.8512(19)	&	7	&	0.121(5)	\\
				&	192	&	28	&	1.0281(12)	&	6	&	1.7546(34)	&	6	&	0.8520(22)	&	6	&	0.122(6)	\\
				&	256	&	28	&	1.0268(14)	&	6	&	1.755(4)	&	6	&	0.8517(27)	&	6	&	0.123(7)	\\
				\colrule
				CC10 &&{\footnotesize$Q>0.1$}&&{\footnotesize$Q>0.3$}&&{\footnotesize$Q>0.3$}&&{\footnotesize$Q>0.8$}& \\
				\colrule
				&	32	 &	5	&	1.0219(18)	&	5	&	1.7630(7)	&	1	&	0.858(6)	&	1	&	0.1200(40)	\\
				&	48	 &	6	&	1.0207(14)	&	5	&	1.7626(6)	&	3	&	0.857(4)	&	3	&	0.1211(30)	\\
				&	64	 &	25	&	1.0180(6)	&	7	&	1.7616(11)	&	7	&	0.8621(9)	&	6	&	0.1243(36)	\\
				&	80		&	28	&	1.0168(5)	&	7	&	1.7598(12)	&	6	&	0.8643(17)	&	7	&	0.1254(25)	\\
				&	96		&	26	&	1.0143(6)	&	7	&	1.7588(12)	&	7	&	0.8651(6)	&	7	&	0.1247(27)	\\
				&	128		&	27	&	1.0129(6)	&	7	&	1.7576(12)	&	7	&	0.8658(6)	&	7	&	0.1245(27)	\\
				&	160		&	27	&	1.0115(6)	&	7	&	1.7563(11)	&	7	&	0.8662(6)	&	7	&	0.1251(27)	\\
				&	192		&	28	&	1.0102(6)	&	7	&	1.7554(11)	&	7	&	0.8666(7)	&	6	&	0.1251(27)	\\
				&	256		&	28	&	1.0099(7)	&	7	&	1.7550(13)	&	7	&	0.8668(8)	&	6	&	0.1251(33)	\\
				\colrule
				CC6-HCPP &&{\footnotesize$Q>0.3$}&&{\footnotesize$Q>0.3$}&&{\footnotesize$Q>0.4$}&&{\footnotesize$Q>0.6$}& \\
				\colrule
				&	32	&	5	&	1.0076(7)	&	2	&	1.7519(23)	&	5	&	0.8731(4)	&	4	&	0.1236(14)	\\
				&	48	&	5	&	1.0068(7)	&	2	&	1.7515(22)	&	5	&	0.8733(4)	&	4	&	0.1240(13)	\\
				&	64	&	24	&	1.0060(5)	&	7	&	1.7598(8)	&	7	&	0.8739(5)	&	6	&	0.1254(19)	\\
				&	80	&	28	&	1.0056(7)	&	7	&	1.7565(9)	&	7	&	0.8738(6)	&	6	&	0.1248(25)	\\
				&	96	&	28	&	1.0054(7)	&	7	&	1.7548(8)	&	7	&	0.8734(6)	&	6	&	0.1240(24)	\\
				&	128	&	28	&	1.0048(7)	&	7	&	1.7541(8)	&	7	&	0.8736(6)	&	6	&	0.1235(24)	\\
				&	160	&	28	&	1.0042(8)	&	7	&	1.7535(8)	&	7	&	0.8739(7)	&	6	&	0.1236(26)	\\
				&	192	&	28	&	1.0030(8)	&	7	&	1.7523(8)	&	7	&	0.8735(7)	&	6	&	0.1253(26)	\\
				&	256	&	26	&	1.0025(9)	&	7	&	1.7518(9)	&	7	&	0.8731(8)	&	5	&	0.1262(30)	\\
			\end{tabular} 
		\end{ruledtabular}
		\caption{List of effective critical exponents for the VD, CC4, CC6, CC10, and CC6 on HCPP lattices. $N_\text{fits}$ denotes the number of fits with a goodness-of-fit parameter $Q$ larger than the value indicated out of $4 \cdot 7 = 28$ fits for $\nu$ and 7 for all other exponents. $L$ denotes the center of the respective fitting window, see Fig.~\ref{fig:weights}. The data is displayed in Figs.~\ref{fig:matrixplot} and \ref{fig:matrixplot_small}.}
		\label{tab:fullresults}
	\end{table*}
	\endgroup

\FloatBarrier

\bibliography{library} 

\end{document}